\documentclass[12pt]{article}

\usepackage[utf8]{inputenc}  
\usepackage{csquotes}        

\usepackage[verbose,
a4paper,         
top=25mm,
bottom=25mm,
left=25mm,
right=25mm]{geometry}

\usepackage{amsmath,amssymb,amsfonts}  
\usepackage{bm}                         
\usepackage{mathtools}                  
\usepackage{amsthm}                      

\usepackage{physics2}                    
\usepackage{braket}                      
\usepackage{siunitx}                      
\usepackage{dsfont}                       
\usepackage{mathrsfs}                     
\usepackage{bbm}                          

\usepackage{graphicx}                     
\usepackage{epstopdf}                      
\usepackage{booktabs}                      
\usepackage{multirow}                      
\usepackage{array}                          
\usepackage{float}                          

\usepackage{xcolor}                         
\usepackage{tikz}                            
\usetikzlibrary{arrows.meta, positioning, calc} 

\usepackage[square,numbers,compress]{natbib}

\usepackage{doi}
\usepackage{xurl}        
\usepackage{breakurl}    

\renewcommand{\doi}[1]{\href{https://doi.org/#1}{https://doi.org/#1}}

\usepackage{authblk}

\usepackage{abstract}

\usepackage{textcomp}      
\usepackage{microtype}     
\usepackage{caption}       
\usepackage{subcaption}    
\usepackage{enumitem}      

\usepackage{hyperref}
\hypersetup{
	colorlinks=true,
	urlcolor=blue,
	citecolor=green!50!black,
	linkcolor=red!50!black,
	pdfauthor={Author(s)},
	pdftitle={Article Title},
	pdfsubject={Physics},
	pdfkeywords={}
}

\begin{document}
	
	\title{Secondary gravitational waves against a strong gravitational wave in the Bianchi VI universe}
	
	\author[1,2]{Konstantin E. Osetrin}

\affil[1]{%
Center for Theoretical Physics,  
Tomsk State Pedagogical University, \newline
Tomsk, 
Russia
}	
\affil[2]{%
International Laboratory for Theoretical Cosmology,\newline 
Tomsk State University of Control Systems and Radioelectronics,\newline 
	Tomsk, 
	Russia
}

\date{}  

\maketitle
	

\begin{abstract}
A proper-time method for constructing models of dynamic gravitational-wave fields is presented. Using the proper-time method, analytical (not numerical) models of secondary gravitational waves are constructed as perturbative solutions of linearized field equations against the background of the exact wave solution of Einstein's equations for the vacuum in the Bianchi VI universe in a privileged wave coordinate system. Relations for the proper time of test particles against the background of a strong gravitational wave are used. The analytical form of the metric components for secondary gravitational waves is found from compatibility conditions for the field equations. From the field equations, an explicit form of ordinary differential equations and their solutions is obtained for functions included in small corrections to the metric for secondary gravitational waves. It is shown that there exists a continuum of gravitational wave parameters for which the perturbative solutions are stable.
\end{abstract}
	
	
\section*{Introduction}\label{sec1}

Relic gravitational waves are an important object for research \cite{MAGGIORE2000283,PhysRevD.75.123518,Caprini2018_163001}, 
since such waves could have a special influence on the formation of the relic electromagnetic background of the universe \cite{PhysRevLett.78.2054}, including the observed anisotropy of the microwave electromagnetic background \cite{Bennett2013}, and could have influenced the formation of initial inhomogeneities and black holes due to tidal accelerations and the creation of sound waves in the relic plasma and relic matter \cite{Saito200916,Saito2010867}. In a number of previous works, well-founded doubts were also expressed in the kinematic interpretation of the existing dipole anisotropy of the relic microwave electromagnetic background of the universe \cite{Secrest_2021,refId0,10.1093/mnras/stad3706}. This provides additional motivation for studying anisotropic models of Bianchi universes and studying gravitational and electromagnetic waves in such models. It has been noted that taking into account the anisotropy of the universe in models of gravitational wave propagation in Bianchi spaces can give a noticeable difference in the observed luminosity of stars and significantly influence the interpretation of modern observational data \cite{LUDWICK2025139717}.

Constructing models of such strong relic gravitational waves requires obtaining exact, rather than perturbative, solutions of nonlinear field equations in Einstein's classical theory of gravity \cite{OBUKHOV2024169816,ObukhovSym15030648} or, even more difficult, for the field equations of modified theories of gravity \cite{Odintsov2007,Odintsov2011,Capozziello2011,Odintsov2017}.
Note that the equality of the speed of gravitational waves and the speed of light established by observations of gravitational waves and electromagnetic radiation from neutron star mergers \cite{Abbott2017PRL161101,Abbott_2017AJL,Abbott_PhysRevX.9.011001} has reduced the number of realistic theoretical models of gravitational waves. Realistic models of gravitational waves can thus be effectively described in privileged wave coordinate systems using wave variables along which the space-time interval vanishes. A number of such exact models of relic gravitational waves were constructed in our papers for the Bianchi universes of types IV, VI, VII and III \cite{Osetrin2022EPJP856,Osetrin2022894,Osetrin325205JPA_2023,Osetrin2023Universe356}. It was shown, in particular, that for the considered exact models of gravitational waves, the tidal accelerations created by them have a transverse orientation to the direction of wave propagation, as well as for perturbative wave models. The presence of exact solutions of the field equations for relic gravitational waves allows us to raise the question of models of secondary gravitational waves against the background of the basic strong gravitational wave and to study complex models of interacting gravitational waves, including secondary gravitational waves. Analytical models for such complex problems have not yet been presented in the literature.

In this paper, we constructed three versions of analytical models for secondary gravitational waves against the background of a strong relic gravitational wave in the Bianchi type VI universe. The possibility of solving a complex system of field equations appeared after we had previously established relations for the proper time of test particles and the retarded time for particle motion and radiation propagation against the background of a strong gravitational wave \cite{OSETRIN2024169619}. The retarded time relation establishes a connection between the coordinates of the world points of radiation emission and radiation observation against the background of an exact solution of the field equations for a strong gravitational wave. In coordinate systems with a time variable, this relation gives the retardation time of radiation propagation against the background of a strong gravitational wave. For secondary gravitational waves propagating at the speed of light, the found relations make it possible to find a solution to the field equations for a complex system of interacting gravitational waves.

The construction of such analytical complex composite models of strong relic gravitational waves and secondary gravitational waves against their background allows for further research into the influence of secondary gravitational waves on the processes that occurred in the early stages of the dynamics of the universe.

Methods for detecting gravitational waves are constantly evolving \cite{PhysRevLett.116.061102,PhysRevX.9.031040,PhysRevX.11.021053}, as are the mathematical methods for studying them
\cite{BlanchetPostNewtonian2024,Christensen2018016903,Osetrin2020Symmetry,Domenech2021398,
	Osetrin1455Sym_2023,van_Remortel_2023,Obukhov10.1142/S0219887825501774,Obukhov10.3390/sym16101385}.

Thus, in the work \cite{Osetrin2024Symmetry1456}, 
exact solutions of the Einstein equations for non-perturbative models of a plane gravitational wave in the presence of pure radiation were found, which demonstrated the possibility of constructing complex exact models in the presence of gravitational waves and radiation.
In the work \cite{OSETRIN2024169619}, an explicit form of the equations for the delay in the propagation of radiation in the exact model of a gravitational wave was found, which opens the possibility of analytically calculating the delay of radiation from stars and galaxies in the presence of a gravitational-wave background and with the possibility of taking into account the anisotropy of space.
In the work \cite{Osetrin2024RPJ1857}, it was shown that the metric of the exact model of a strong plane gravitational wave is incompatible with a source in the form of dust matter in the Einstein equations and the wave metric in the presence of matter requires generalization.

The study of the trajectories of light rays in a gravitational field is an important mathematical and physical element of research in the interests of observational astronomy \cite{LandauEng1}. Such problems are usually solved on the basis of perturbative and numerical methods \cite{MUKHANOV1992203,Ma19957}.
Analytical models of gravitational waves provide both new approaches and calculation methods, and develop a mathematical basis for perturbative and numerical methods.

Methods for detecting gravitational waves are constantly evolving \cite{PhysRevLett.116.061102,PhysRevX.9.031040,PhysRevX.11.021053}, as are the mathematical methods for studying them
\cite{Christensen2018016903,Osetrin2020Symmetry,Domenech2021398,
	Osetrin1455Sym_2023,van_Remortel_2023,Obukhov10.1142/S0219887825501774,Obukhov10.3390/sym16101385}.
Based on new mathematical approaches, a number of exact mathematical models of gravitational waves and, in particular, relic gravitational waves in various cosmological models have been constructed \cite{OsetrinHomog2006,Osetrin2022EPJP856,Osetrin2022894,Osetrin325205JPA_2023}. A number of exact cosmological models with an electromagnetic field have also been constructed, which allow integrating the equations of charge motion \cite{ObukhovSym15030648,ObukhovSym14122595,ObukhovUniverse8040245,Obukhov10.1063/5.0158054}, which is important for studying relic plasma.

Recently published results of long-term observations of pulsar signal delays \cite{AandArefId0,Reardon_2023,Xu_2023} allow us to judge the presence and characteristics of the gravitational-wave background of the universe, the level of influence of this background on other observational data due to the effect of ''the retarded time of radiation'' in gravitational waves.

\section{Exact model of gravitational wave \\in the Bianchi type VI universe}\label{sec2}

The model of a strong gravitational wave, whose metric in a privileged wave coordinate system depends on the wave variable, was developed by a number of researchers and is given in the textbook by Landau and Lifshitz \cite{LandauEng1}. This wave model belongs to spaces that allow integration of the equations of motion of test particles in the Hamilton-Jacobi formalism by the method of separation of variables and, thus, belongs to the classes of Stackel spaces
\cite{Stackel1897145,Shapovalov1978I, Shapovalov1978II,Shapovalov1979} and Shapovalov spaces \cite{Osetrin2020Symmetry}. Based on this exact model, a number of exact wave solutions for Bianchi spaces were constructed, which describe homogeneous but non-isotropic universes, i.e. these models can also serve to describe relict gravitational waves in the universe at the early stages of its dynamics. An exact model of a gravitational wave in the Bianchi type VI universe was obtained and investigated in a number of our papers, where exact solutions for the trajectories of test particles and light were found, solutions to the equations of deviation of geodesics were found, analytical expressions for tidal accelerations were found, and relations for the delay of radiation propagating against the background of a gravitational wave were found 
\cite{
Osetrin325205JPA_2023,OSETRIN2024169619,OsetrinHomog2006}.

In a privileged wave coordinate system with wave variables \(x^0\) and \(x^1\), along which the space-time interval vanishes, the wave metric for a homogeneous anisotropic Bianchi type VI spacetime can be represented in the following form \cite{Osetrin325205JPA_2023}:
\begin{equation}
	ds^2=2dx^0dx^1
	-
	\frac{\bigl(x^0 \bigr)^{2\mu}{dx^2}^2
		+2\cos(\phi)\, \bigl( x^0 \bigr)^{\mu+\nu}{dx^2}{dx^3}
		+\bigl( x^0 \bigr)^{2\nu}{dx^3}^2}{\sin^2(\phi)}
	\label{MetricVI}
	,\end{equation}
\begin{equation}
	0<\phi<\pi
	,\end{equation}
where the constants $\phi$, $\mu$ and $\nu$ are independent parameters of the wave model.

This spacetime model admits a group of spatial homogeneity motions with Killing vectors $X_{(1)}$, $X_{(2)}$ and $X_{(3)}$, which in the privileged wave coordinate system have the simple form:
\begin{equation}
	X^\alpha_{(1)}=\bigl(0,0,1,0\bigr),
	\quad
	X^\alpha_{(2)}=\bigl(0,0,0,1\bigr),
	\quad
	X^\alpha_{(3)}=\bigl(-x^0,x^1,\mu\, x^2,\nu\, x^3\bigr)
	.\end{equation}
The commutation relations for the Killing vectors of the spatial homogeneity group take the form:
\begin{equation}
	\left[X_{(1)},X_{(2)}\right]=0
	,\qquad
	\left[X_{(1)},X_{(3)}\right]=\mu X_{(1)}
	,\qquad
	\left[X_{(2)},X_{(3)}\right]=\nu X_{(2)}
	,\end{equation}
which defines a homogeneous anisotropic Bianchi space of type VI.

It is easy to see that this wave space admits a covariantly constant vector
\begin{equation}
	\nabla_\alpha K_\beta=0
	\quad
	\to
	\quad
	K_\alpha=\bigl( 1,0,0,0 \bigr)
	,\quad
	K^\alpha=\bigl( 0,1,0,0 \bigr)
	,\end{equation}
which specifies the direction of propagation of the gravitational wave.

The Riemann curvature tensor in the privileged coordinate system with the wave variable \(x^0\) has the following three non-zero components:
\begin{equation}
	{ R }_{0202} = \frac{
		(\mu-\nu)^2+
		\sin^2(\phi) \Bigl(
		4\mu(1-\mu)-(\mu-\nu)^2
		\Bigr)
	}{4 \sin^4(\phi)}
	\,\bigl( x^0 \bigr)^{2 \mu -2} 
\end{equation}
\begin{equation}
	{ R }_{0302} = \frac{{\cos(\phi)} 
		\Bigl(
		\sin^2(\phi)(\mu+\nu)(\mu+\nu-2)-(\mu-\nu)^2
		\Bigr)
	}{4 \sin^4(\phi)}
	\,\bigl( x^0 \bigr)^{\mu +\nu -2}
\end{equation}
\begin{equation}
	{ R }_{0303} = \frac{
		(\mu-\nu)^2
		+
		\sin^2(\phi)\Bigl(
		4\nu(1-\nu)-(\mu-\nu)^2
		\Bigr)
	}{4 \sin^4(\phi)}
	\,
	\bigl( x^0 \bigr)^{2 \nu -2} 
\end{equation}
The Ricci tensor ${ R }_{\alpha\beta}={ R }^\gamma{}_{\alpha\gamma\beta}$ in the privileged wave coordinate system has one nonzero component:
\begin{equation}
	{ R }_{00} = -\frac{
		(\mu-\nu)^2
		+
		\sin^2(\phi)(\mu+\nu)(\mu+\nu-2)
	}{2 \sin^4(\phi)\, \bigl( x^0 \bigr)^2}
	.
\end{equation}
The scalar curvature of ${ R }$ is zero.

The Weyl conformal curvature tensor has the following nonzero components:
\begin{equation}
	{\rm C}_{0202} = -\frac{(\mu -\nu ) \left[
		\sin^2(\phi)(2\mu-1)-\mu+\nu
		\right]}{2 \sin^4(\phi)}
	\,
	\bigl( x^0 \bigr)^{2 \mu -2} 
	,\end{equation}
\begin{equation}
	{\rm C}_{0302} = -\frac{\cos(\phi)(\mu -\nu )^2 }{2 \sin^4(\phi)}
	\,
	\bigl( x^0 \bigr)^{\mu +\nu -2}
	,\end{equation}
\begin{equation}
	{\rm C}_{0303} = \frac{(\mu -\nu ) 
		\left[
		\sin^2(\phi)(2\nu-1)+\mu-\nu
		\right]
	}{2 \sin^4(\phi)}
	\,
	\bigl( x^0 \bigr)^{2 \nu -2}
	.\end{equation}
Thus, when the parameters are equal $\mu=\nu$, the model degenerates and the space becomes conformally flat (or flat).

The metric gravitational wave model (\ref{MetricVI}) gives an exact solution to the Einstein vacuum equations (the cosmological constant for this model vanishes) under a single condition on the original three parameters of the model (\(\mu\), \(\nu\) and \(\phi\)):
\begin{equation}
	(\mu-\nu)^2
	+
	(\mu+\nu)(\mu+\nu-2)\sin^2(\phi)=0
	\label{MuNuPhi}
	.\end{equation}
Thus, in the exact gravitational wave model for the classical Einstein theory of gravity in a homogeneous anisotropic Bianchi space of type VI, only two independent parameters remain, which we can fix by introducing two independent angular parameters through the following relations:
\begin{equation}
	{\mu}=\frac{1}{2}\,
	\left(
	1+\cos{\theta} + \sin{\phi} \sin{\theta}
	\right)
	\label{MetricA}
	,\end{equation}
\begin{equation}
	{\nu}=\frac{1}{2}\,
	\left(
	1+\cos{\theta} - \sin{\phi} \sin{\theta}
	\right)
	\label{MetricB}
	,\end{equation}
\begin{equation}
	0<\phi <\pi
	,\qquad
	0<\theta <\pi
	\label{MetricC}
	,\end{equation}
where instead of three parameters $\mu$, $\nu$ and $\phi$ there remain only two independent angular parameters -- \(\phi\) and \(\theta\).

Note that the Riemann curvature tensor in this case can identically vanish (degeneration of the model) when two conditions on the parameters $\phi=\pi/2$ and $\theta=\pi/2$ are simultaneously satisfied, when $\mu=\nu=1$.

\section{Proper time of a particle freely moving \\in a strong gravitational wave}\label{sec3}

In \cite{Osetrin325205JPA_2023} the equations of the trajectories of test particles were integrated in the Hamilton-Jacobi formalism for the exact model of a gravitational wave in the Bianchi space of type VI. The results obtained allow us to find relations for the analytical form of the proper time of a test particle in a gravitational wave. For our model, this leads us to three possible cases. These three different cases are determined by the following relations for the parameters of the gravitational wave (it is additionally assumed that the non-degeneracy condition of the model \(\mu\ne\nu\) is satisfied):
\begin{enumerate}
	\item
	The parameter \(\mu\) or the parameter \(\nu\) are equal to \(1/2\);
	\item
	The sum of the parameters \(\mu+\nu\) is equal to 1;
	\label{item2}
	\item
	Other permissible values of the parameters \(\mu\) and \(\nu\) that do not apply to the first two cases.
\end{enumerate}
The first case, due to the symmetry of the gravitational-wave model, when replacing the coordinates \(x^2\) and \(x^3\) or, what is the same, replacing the parameters \(\mu\) and \(\nu\) is equivalent to one case. Therefore, in the first case, for definiteness, we will choose the option \(\mu=1/2\) and \(\nu\ne 1/2\), then for the time \(\tau\) according to the clock of the test particle in the gravitational wave we will find the following expression for \(\tau\) through the coordinates \(x^\alpha\):
\[
\tau^2=x^0\Biggl(
2 x^1+
(2\nu-1)\,
\frac{
	\left( x^2\right)^2 
	-4\cos(\phi)\left({x^0}\right)^{\nu-1/2} x^2  x^3
}{4 \cos ^2(\phi )+(2 \nu -1) \log ({x^0})}
\]
\begin{equation}
	-
	\frac{
		(2\nu-1)^2 \log(x^0)\, \bigl( x^0 \bigr)^{2\nu-1}\left( x^3\right)^2
	}{4 \cos ^2(\phi )+(2 \nu -1) \log ({x^0})}
	\Biggr)
	\label{tauA}
	.\end{equation}

In the second special case, for \(\mu=1-\nu\) and \(\nu\ne 1/2\) we obtain the following relationship:
\[
\tau^2=
2
x^0
\Biggl(
x^1
+
4\,
\frac{
	(2 \nu -1) \bigl( x^0 \bigr)^{1-2 \nu }\left( x^2\right)^2 
	+2 \sin ^2(\phi ) \cos (\phi ) \log ({x^0})\, x^2  x^3
}{\left(1-\cos (4 \phi )\right) \log ^2({x^0})
	+8
}
\]
\begin{equation}
	-\frac{
		4\, (2 \nu -1) \bigl( x^0 \bigr)^{2 \nu -1}\left( x^3\right)^2
	}{\left(1-\cos (4 \phi )\right) \log ^2({x^0})
		+8}
	\Biggr)
	\label{tauB}
	.\end{equation}

In the third (most general) case for \(\mu\ne\nu\), \(\mu+\nu\ne 1\) and \(\mu,\nu\ne 1/2\) we find:
\[
\tau^2=
x^0\Biggl(
2 x^1
-
\frac{
	(2 \mu  -1) (\mu  +\nu  -1)^2(\cos (2 \phi )-3) \csc ^2(\phi ) \left( x^0 \right)^{2 \mu  -1} \left( x^2\right)^2 
}{2(\mu+\nu)-4\mu\nu -3+ (2 \mu  -1) (2 \nu  -1) \cos (2 \phi )}
\]
\[
-
\frac{
	(2 \mu  -1) (2 \nu  -1) (\mu  +\nu  -1) (\cos (2 \phi )-3) \cot (\phi ) \csc (\phi) \left( x^0 \right)^{\mu  +\nu  -1} x^2  x^3
}{2(\mu+\nu)-4\mu\nu -3+ (2 \mu  -1) (2 \nu  -1) \cos (2 \phi )}
\]
\begin{equation}
	+
	\frac{
		(2 \nu  -1) (\mu  +\nu  -1)^2 (\cos(2 \phi )-3) \csc ^2(\phi )
		\left( x^0 \right)^{2 \nu  -1} 
		\left( x^3\right)^2 
	}{2(\mu+\nu)-4\mu\nu -3+ (2 \mu  -1) (2 \nu  -1) \cos (2 \phi )}
	\Biggr)
	\label{tauC}
	.\end{equation}

As was shown in \cite{Osetrin325205JPA_2023}, the proper time \(\tau\) of a test particle in a gravitational wave can be chosen as the common time of a new synchronous reference frame \cite{LandauEng1} with separation of time and space variables and with the reference frame tied to this test particle, freely ''falling'' in a gravitational wave.

\section{Perturbative model of secondary gravitational waves}\label{sec4}

The existence of an exact wave solution of the Einstein equations in vacuum allows us to formulate the problem of finding a perturbative model of weak secondary gravitational waves against the background of a "strong" exact solution. Based on the wave nature of gravitational waves propagating at the speed of light, it is rational to use a privileged wave coordinate system with conjugate zero variables \(x^0\) and \(x^1\), along which the space-time interval vanishes, i.e. the metric components \(g_{00}\) and \(g_{11}\) in such a wave coordinate system vanish, which corresponds to the form of the background metric of a strong gravitational wave (\ref{MetricVI}). The solution of the equations of the trajectory of test particles in a strong gravitational wave  (\ref{MetricVI}) in the paper \cite{Osetrin325205JPA_2023} allows us to introduce a synchronous time variable \(\tau\), which corresponds to the clock time of an observer freely moving in a strong gravitational wave.
Therefore, we will seek a perturbative solution of the field equations for the secondary gravitational wave in the form of small corrections that depend both on the non-ignorable wave variable \(x^0\) (the metric of the background exact wave solution of the Einstein equations depends on it) and on the synchronous time \(\tau\). The variables \(x^2\) and \(x^3\) of the privileged wave coordinate system enter the time function \(\tau\) quadratically, and the variable \(x^1\), along which the strong gravitational wave propagates, enters linearly. The dependence of the time variable \(\tau\) on the wave variable \(x^0\) is complex and is determined by the metric of the background strong gravitational wave.

Thus, the perturbative version of the secondary gravitational wave metric \({\tilde g}^{\alpha\beta}\) will be sought in the privileged wave coordinate system described above in the following form:
\begin{equation}
	{\tilde g}^{\alpha\beta}=g^{\alpha\beta}(x^0)+\epsilon\, \Omega^{\alpha\beta}\bigl(x^0,\tau(x^0,x^1,x^2,x^3)\bigr)
	,\end{equation}
where \(g^{\alpha\beta}(x^0)\) is the background strong gravitational wave metric, \(\epsilon\) is the dimensionless smallness parameter (\(\epsilon\ll 1\)).

Accordingly, the metric with subscripts in the linear approximation in \(\epsilon\) (\(\epsilon\ll 1\)) will take the form:
\begin{equation}
	{\tilde g}_{\alpha\beta}=g_{\alpha\beta}(x^0)-\epsilon\, \Omega^{\alpha\beta}\bigl(x^0,\tau(x^0,x^1,x^2,x^3)\bigr)
	.\end{equation}

The Einstein vacuum equations linearized with respect to the smallness parameter~\(\epsilon\) (the cosmological constant vanishes) are too cumbersome to be presented here.

The study of the compatibility conditions with respect to the variables \(x^2\) and~\(x^3\) for the obtained linearized field equations after fairly cumbersome intermediate calculations yields the following simple structure of the dependence of the metric components \({\tilde g}_{\alpha\beta}\) on the time function \(\tau(x^0,x^1,x^2,x^3)\) and the wave variable \(x^0\):
\begin{equation}
	\Omega^{12} = B_{12}(x^0)\, \tau^2(x^0,x^1,x^2,x^3)+A_{12}(x^0)
	,\end{equation}
\begin{equation}
	\Omega^{13} = B_{13}(x^0)\, \tau^2(x^0,x^1,x^2,x^3)+A_{13}(x^0)
	,\end{equation}
\begin{equation}
	\Omega^{00} = 0
	,\qquad
	\Omega^{11} = 0
	,\end{equation}
\begin{equation}
	\Omega^{01} = A_{01}(x^0)
	,\qquad
	\Omega^{02} = A_{02}(x^0)
	,\qquad
	\Omega^{03} = A_{03}(x^0)
	,\end{equation}
\begin{equation}
	\Omega^{22} = A_{22}(x^0)
	,\qquad
	\Omega^{23} = A_{23}(x^0)
	,\qquad
	\Omega^{33} = A_{33}(x^0)
	,\end{equation}
where \(A_{\alpha\beta}(x^0)\) and \(B_{\alpha\beta}(x^0)\) are functions of the wave variable \(x^0\) only.
Note that functions \(A_{01}(x^0)\), \(A_{12}(x^0)\) and \(A_{13}(x^0)\) can be converted to zero by coordinate transformations.

The form of the linearized vacuum field equations for the functions \(A_{\alpha\beta}(x^0)\) and \(B_{\alpha\beta}(x^0)\) depends on the range of background wave parameters \(\mu\), \(\nu\), and \(\phi\) (or independent angular parameters \(\phi\) and \(\theta\)), for which there are three different versions of the time function (\ref{tauA})-(\ref{tauC}).

In this section we will consider the most general case for \(\mu\ne\nu\), \(\mu+\nu\ne 1\) and \(\mu,\nu\ne 1/2\), when the time variable \(\tau\) takes the form (\ref{tauC}). In this case, for a more compact notation of the equations, it is convenient to use independent angular parameters \(\phi\) and \(\theta\) (\ref{MetricA})-(\ref{MetricC}) instead of the parameters \(\mu\), \(\nu\) and \(\phi\) related by the relation (\ref{MuNuPhi}).
Then, after taking into account the compatibility conditions, from the field equations we obtain a system of coupled linear homogeneous differential equations with variable coefficients for the functions \(B_{12}(x^0)\) and \(B_{13}(x^0)\):
\[
B_{12}' = 
\frac{ \cot (\phi ) \sin (\theta )\, \bigl( x^0 \bigr)^{-\sin (\phi ) \sin (\theta )-1} 
	B_{13}
}{\cos ^2(\phi ) \sin ^2(\theta )+\cos ^2(\theta )}
\biggl(
2 \cos ^2(\theta )
+
\cos ^2(\phi ) \sin ^2(\theta )
\]
\[
\mbox{}
-\sin (\phi ) \sin (\theta ) \cos (\theta )
\biggr)
-\frac{
	B_{12} 
}{{x^0} \left(\cos ^2(\phi ) \sin ^2(\theta )+\cos ^2(\theta )\right)}
\biggl(
2 \cos ^3(\theta )
\]
\[
\mbox{}
+2 \cos ^2(\phi ) \sin ^2(\theta ) \cos (\theta )
+2 \csc (\phi ) \cos ^2(\theta ) \bigl(\sin (\phi )+\sin (\theta )\bigr)
\]
\begin{equation}
	+\cos (\phi ) \cot (\phi ) \sin ^2(\theta ) 
	\bigl(
	2 \sin (\phi )+\sin (\theta )
	\bigr)
	\biggr)
	\label{B12eqA}
	,\end{equation}
\[
B_{13}' = \frac{B_{13} 
}{{x^0} \left(\cos ^2(\phi ) \sin ^2(\theta )+\cos ^2(\theta )\right)}
\biggl(
-2 \cos ^2(\phi ) \sin ^2(\theta ) \cos (\theta )
-2 \cos ^3(\theta )
\]
\[
\mbox{}
+2 \cos ^2(\theta ) \bigl(\csc (\phi ) \sin (\theta )-1\bigr)
+\cos (\phi ) \sin ^2(\theta ) \Bigl(\cot (\phi ) \sin (\theta )-2 \cos (\phi )\Bigr)
\biggr)
\]
\[
\mbox{}
-
\frac{
	B_{12}\cos (\phi ) \sin (\theta ) \bigl( x^0 \bigr)^{\sin (\phi ) \sin (\theta )-1} 
}{ \cos ^2(\phi ) \sin ^2(\theta )+\cos ^2(\theta )}
\biggl(
\cos (\phi ) \cot (\phi ) \sin ^2(\theta )
\]
\begin{equation}
	\mbox{}
	+
	2\, \frac{ \cos ^2(\theta )}{\sin(\phi )}+ \sin (\theta ) \cos (\theta )
	\biggr)
	\label{B13eqA}
	.\end{equation}
Here the prime denotes the derivative with respect to the wave variable \(x^0\).

For the parameter value \(\phi=\pi/2\), the equations in the system (\ref{B12eqA})-(\ref{B13eqA}) "disconnect" and become independent due to the presence of the factors \(\cot (\phi )\) and \(\cos (\phi )\).
The system of coupled equations (\ref{B12eqA})-(\ref{B13eqA}) for the parameter values \(\phi\ne\pi/2\), however, also allows one to obtain separate independent equations
for the functions \(B_{12}(x^0)\) and \(B_{13}(x^0)\) by repeated differentiation
of the equations (\ref{B12eqA})-(\ref{B13eqA}).

As a result of differentiating the equation (\ref{B12eqA}) and substituting the first derivatives, we obtain for the function \(B_{12}(x^0)\) an independent linear homogeneous differential equation of the second order with variable coefficients:
\[
B_{12}'' = 
-\frac{ \bigl(\sin (\phi ) \sin (\theta )+4 \cos (\theta )+5\bigr)B_{12}'}{{x^0}}
\]
\[
\mbox{}
-\frac{ \cos ^2\left({\theta }/{2}\right) B_{12}
}{8 \bigl( x^0 \bigr)^2 \left(\cos ^2(\phi ) \sin ^2(\theta )+\cos ^2(\theta )\right)}
\biggl(
3 \cos (\phi -3 \theta )-4 \cos (2 \phi -3 \theta )+7 \cos (\phi -\theta )
\]
\[
\mbox{}
-8 \cos (2 (\phi -\theta ))
-\cos (3 (\phi -\theta ))
+4 \cos (2 \phi -\theta )+3 \cos (3 \phi -\theta )-7 \cos (\phi +\theta )
\]
\[
\mbox{}
-8 \cos (2 (\phi +\theta ))+\cos (3 (\phi +\theta ))+4 \cos (2 \phi +\theta )
-3 \cos (3 \phi +\theta )
-3 \cos (\phi +3 \theta )
\]
\begin{equation}
	\mbox{}
	-4 \cos (2 \phi +3 \theta )+16 \cos (2 \phi )+80 \cos (\theta )+16 \cos (3 \theta )+32
	\biggr)
	\label{B12eqB}
	.\end{equation}

For the function \(B_{13}(x^0)\) differentiating the equation (\ref{B13eqA}) we obtain the following second-order differential equation:
\[
B_{13}'' = \frac{ \bigl(\sin (\phi ) \sin (\theta )-4 \cos (\theta )-5\bigr)B_{13}'}{{x^0}}
\]
\[
\mbox{}
+\frac{\cos ^2\left({\theta }/{2}\right) B_{13} 
}{\bigl( x^0 \bigr)^2 (\cos (2 (\phi -\theta ))+\cos (2 (\phi +\theta ))-2 \cos (2 \phi )-2 \cos (2 \theta )-6)}
\times
\]
\[
\times
\biggl(
32
+
16 \cos (3 \theta )
-3 \cos (\phi -3 \theta )-
4 \cos (2 \phi -3 \theta )
-7 \cos (\phi -\theta )
\]
\[
\mbox{}
+\cos (3 (\phi -\theta ))
+4 \cos (2 \phi -\theta )-3 \cos (3 \phi -\theta )+7 \cos (\phi +\theta )
+80 \cos (\theta )
\]
\[
-8 \cos (2 (\phi +\theta ))
-8 \cos (2 (\phi -\theta ))
-\cos (3 (\phi +\theta ))+4 \cos (2 \phi +\theta )
\]
\begin{equation}
	\mbox{}
	+3 \cos (3 \phi +\theta )
	+3 \cos (\phi +3 \theta )
	-4 \cos (2 \phi +3 \theta )+16 \cos (2 \phi )
	\biggr)
	\label{B13eqB}
	.\end{equation}

The solution of linear differential equations with variable coefficients (\ref{B12eqB})-(\ref{B13eqB}) can be found in the form of power functions \((x^0)^{a}\), where the power exponents \({a}\) will be determined through the angular parameters of the background wave \(\phi\) and \(\theta\).

The general solution to Eq.(\ref{B12eqB}) is as follows
\begin{equation}
B_{12}(x^0)=c_1\, {(x^0)}^{a_1}+c_2\, {(x^0)}^{a_2}
,\qquad
c_1, c_2\mbox{ -- const}
,\end{equation}
where \(a_1\) and  \(a_2\)   satisfies the quadratic equation:
\[
0=a^2+\left(\sin (\phi ) \sin (\theta )+4 \cos (\theta )+4\right)
\,a
\]
\[
\mbox{}
+\frac{ \cos ^2\left({\theta }/{2}\right) 
}{8  \left(\cos ^2(\phi ) \sin ^2(\theta )+\cos ^2(\theta )\right)}
\biggl(
3 \cos (\phi -3 \theta )-4 \cos (2 \phi -3 \theta )+7 \cos (\phi -\theta )
\]
\[
\mbox{}
-8 \cos (2 (\phi -\theta ))
-\cos (3 (\phi -\theta ))
+4 \cos (2 \phi -\theta )+3 \cos (3 \phi -\theta )-7 \cos (\phi +\theta )
\]
\[
\mbox{}
-8 \cos (2 (\phi +\theta ))+\cos (3 (\phi +\theta ))+4 \cos (2 \phi +\theta )
-3 \cos (3 \phi +\theta )
-3 \cos (\phi +3 \theta )
\]
\begin{equation}
	\mbox{}
	-4 \cos (2 \phi +3 \theta )+16 \cos (2 \phi )+80 \cos (\theta )+16 \cos (3 \theta )+32
	\biggr)
	.\end{equation}
The general solution to Eq.(\ref{B13eqB}) is as follows
\begin{equation}
	B_{13}(x^0)=c_3\, {(x^0)}^{b_1}+c_4\, {(x^0)}^{b_2}
	,\qquad
	c_3, c_4\mbox{ -- const}
	,\end{equation}
where \(b_1\) and \(b_2\)  satisfies the following quadratic equation:
\[
0=b^2
+
\bigl(-\sin (\phi ) \sin (\theta )+4 \cos (\theta )+4\bigr)
\,b
\]
\[
\mbox{}
-\frac{\cos ^2\left({\theta }/{2}\right) 
}{\cos (2 (\phi -\theta ))+\cos (2 (\phi +\theta ))-2 \cos (2 \phi )-2 \cos (2 \theta )-6}
\times
\]
\[
\times
\biggl(
16 \cos (3 \theta )
-3 \cos (\phi -3 \theta )-
4 \cos (2 \phi -3 \theta )
-7 \cos (\phi -\theta )
\]
\[
\mbox{}
+\cos (3 (\phi -\theta ))
+4 \cos (2 \phi -\theta )-3 \cos (3 \phi -\theta )+7 \cos (\phi +\theta )
+80 \cos (\theta )
\]
\[
-8 \cos (2 (\phi +\theta ))
-8 \cos (2 (\phi -\theta ))
-\cos (3 (\phi +\theta ))+4 \cos (2 \phi +\theta )
\]
\begin{equation}
	\mbox{}
	+3 \cos (3 \phi +\theta )
	+3 \cos (\phi +3 \theta )
	-4 \cos (2 \phi +3 \theta )+16 \cos (2 \phi )
	+32
	\biggr)
	.\end{equation}

From the remaining field equations, a system of coupled linear inhomogeneous differential equations with variable coefficients on the functions \(A_{02}(x^0)\) and \(A_{03}(x^0)\) additionally arises:
\[
A_{02}'' = - \bigl(
\csc (\phi ) \sin (\theta )+2 \cos (\theta )+2
\bigr)
\,\frac{A_{02}'}{{x^0}}
\]
\begin{equation}
	\mbox{}
	+\cot (\phi ) \sin (\theta )\, \bigl( x^0 \bigr)^{-\sin (\phi ) \sin (\theta )-1} A_{03}'
	+2\bigl(1- \cos (\theta )\bigr) B_{12} 
	\label{A02eqA}
	,\end{equation}

\[
A_{03}'' = -\cot (\phi ) \sin (\theta ) \, A_{02}' \,\bigl( x^0 \bigr)^{\sin (\phi ) \sin (\theta )-1}
\]
\begin{equation}
	\mbox{}
	+ 
	\bigl(
	\csc (\phi ) \sin (\theta )-2 \cos (\theta )-2
	\bigr)\, \frac{A_{03}'}{{x^0}}
	+2\bigl(1- \cos (\theta )\bigr) B_{13} 
	\label{A03eqA}
	.\end{equation}

Note that to integrate the system of equations (\ref{A02eqA})-(\ref{A03eqA}), it is necessary to first integrate the equations (\ref{B12eqA})-(\ref{B13eqB}) and obtain solutions for \(B_{12}(x^0)\) and \(B_{13}(x^0)\). In addition, for the parameter value \(\phi=\pi/2\), the equations in the system (\ref{A02eqA})-(\ref{A03eqA}) "disconnect" and become independent due to the presence of the factor \(\cot (\phi )\).

From the system of equations (\ref{A02eqA})-(\ref{A03eqA}), we can also obtain separate differential equations
for \(A_{02}(x^0)\) and \(A_{03}(x^0)\) by additional differentiation and repeated use of equations (\ref{A02eqA})-(\ref{A03eqA}) and equations (\ref{B12eqA})-(\ref{B13eqA}).

As a result, for \(A_{02}(x^0)\) we obtain the following linear nonhomogeneous differential equation with variable coefficients of the third order:
\[
A_{02}''' = 
-\bigl(\sin (\phi ) \sin (\theta )+4 \cos (\theta )+5\bigr)\,\frac{A_{02}''}{{x^0}}
\]
\[
\mbox{}
-8 \cos ^3\left({\theta }/{2}\right) 
\bigl(
\sin (\phi ) \sin \left({\theta }/{2}\right)+2 \cos \left({\theta }/{2}\right)
\bigr)
\,\frac{A_{02}'
}{\bigl( x^0 \bigr)^2}
\]
\[
\mbox{}
+
\frac{ \csc (\phi ) \sin ^2\left({\theta }/{2}\right)\, B_{12}
}{16 {x^0} \left(\cos ^2(\phi ) \sin ^2(\theta )+\cos ^2(\theta )\right)}
\biggl(
12 \sin (\phi -2 \theta )
-4 \sin (2 \phi -3 \theta )-\sin (4 \phi -3 \theta )
\]
\[
\mbox{}
-4 \sin (3 \phi -2 \theta )
+28 \sin (2 \phi -\theta )
+3 \sin (4 \phi -\theta )
-28 \sin (2 \phi +\theta )-3 \sin (4 \phi +\theta )
\]
\[
\mbox{}
+12 \sin (\phi +2 \theta )-4 \sin (3 \phi +2 \theta )
+4 \sin (2 \phi +3 \theta )
+\sin (4 \phi +3 \theta )
+40 \sin (\phi )
+8 \sin (3 \phi )
\]
\[
\mbox{}
-82 \sin (\theta )-26 \sin (3 \theta )
\biggr)
\]
\[
\mbox{}
+
\frac{2  \cot (\phi ) \sin (\theta ) \bigl(1-\cos (\theta )\bigr)\, \bigl( x^0 \bigr)^{-\sin (\phi ) \sin (\theta )-1}
	B_{13}
}{\cos ^2(\phi ) \sin ^2(\theta )+\cos ^2(\theta )}
\times
\]
\begin{equation}
	\times
	\biggl(
	2 \cos ^2(\phi ) \sin ^2(\theta )
	-\sin (\phi ) \sin (\theta ) \cos (\theta )+3 \cos ^2(\theta )
	\biggr)
	\label{A02eqB}
	.\end{equation}

Similarly, for \(A_{03}(x^0)\) we obtain the following ordinary differential equation (linear non-homogeneous differential equation with variable coefficients of the third order):
\[
A_{03}''' =\bigl(\sin (\phi ) \sin (\theta )-4 \cos (\theta )-5\bigr)\, \frac{A_{03}'' }{{x^0}}
\]
\[
\mbox{}
-8 \cos ^3\left({\theta }/{2}\right)
\bigl(2 \cos \left({\theta }/{2}\right)-\sin (\phi ) \sin \left({\theta }/{2}\right) \bigr)
\, 
\frac{ A_{03}'}{\bigl( x^0 \bigr)^2}
\]
\[
\mbox{}
+\frac{2  \cot (\phi ) \sin (\theta ) (\cos (\theta )-1) \bigl( x^0 \bigr)^{\sin (\phi ) \sin (\theta )-1} B_{12}
}{\cos ^2(\phi ) \sin ^2(\theta )+\cos ^2(\theta )}
\times
\]
\[
\times
\biggl(
2 \cos ^2(\phi ) \sin ^2(\theta )
+\sin (\phi ) \sin (\theta ) \cos (\theta )+3 \cos ^2(\theta )
\biggr)
\]
\[
\mbox{}
+\frac{ \csc (\phi ) \sec ^2\left({\theta }/{2}\right) \, B_{13}
}{64 {x^0} \left(\cos ^2(\phi )+\cot ^2(\theta )\right)}
\, 
\biggl(
4 \sin (2 \phi -3 \theta )+\sin (4 \phi -3 \theta )+12 \sin (\phi -2 \theta )
\]
\[
\mbox{}
-4 \sin (3 \phi -2 \theta )
-28 \sin (2 \phi -\theta )-3 \sin (4 \phi -\theta )+28 \sin (2 \phi +\theta )+3 \sin (4 \phi +\theta )
\]
\[
\mbox{}
+12 \sin (\phi +2 \theta )-4 \sin (3 \phi +2 \theta )
-4 \sin (2 \phi +3 \theta )-\sin (4 \phi +3 \theta )+40 \sin (\phi )
\]
\begin{equation}
	\mbox{}
	+8 \sin (3 \phi )+82 \sin (\theta )+26 \sin (3 \theta )
	\biggr)
	\label{A03eqB}
	.\end{equation}

It is evident that the solution of differential equations (\ref{A02eqB})-(\ref{A03eqB}) can also be found in the form of power functions \((x^0)^{c}\), where the power exponents \({c}\) will be determined through the angular parameters of the background wave \(\phi\) and \(\theta\).

For the functions \(A_{01}(x^0)\), \(A_{22}(x^0)\), \(A_{23}(x^0)\) and \(A_{33}(x^0)\) remaining undefined in the metric of the secondary gravitational wave, only one equation remains from the field equations, connecting these functions, of the following form:
\[
0=
-8 \left(1+\cos (\theta )\right) 
\bigl( x^0 \bigr)^{\cos (\phi ) \cos (\theta )}
A_{01}' 
+4 \csc ^2(\phi ) 
\bigl( x^0 \bigr)^{\cos (\phi -\theta )+\cos (\theta )+2}
A_{22}'' 
\]
\[
\mbox{}
+12 \csc (\phi ) 
\Bigl(\csc (\phi ) \bigl(1+\cos (\theta )\bigr)+\sin (\theta )
\Bigr)
\bigl( x^0 \bigr)^{\cos (\phi -\theta )+\cos (\theta )+1}
A_{22}'
\]
\[
\mbox{}
+
\bigl( x^0 \bigr)^{\cos (\phi -\theta )+\cos (\theta )}
A_{22} \,
\biggl(
\cot ^2(\phi ) \cos ^2(\theta )+12 \csc ^2(\phi ) \cos (\theta )
-2 \cot ^2(\phi ) \sin ^2(\theta )
\]
\[
\mbox{}
+\frac{1}{2} \csc ^2(\phi ) (\cos (2 \theta )+21)
+4 \csc (\phi ) \sin (\theta ) (4 \cos (\theta )+3)+\sin ^2(\theta )+1
\biggr)
\]
\[
\mbox{}
+4 \csc ^2(\phi ) 
\bigl( x^0 \bigr)^{\cos (\phi +\theta )+\cos (\theta )+2}
A_{33}'' 
\]
\[
\mbox{}
+12 \csc (\phi ) 
\Bigl(
\csc (\phi ) \bigl(1+\cos (\theta )\bigr)-\sin (\theta )
\Bigr)
\,
\bigl( x^0 \bigr)^{\cos (\phi +\theta )+\cos (\theta )+1} 
A_{33}'
\]
\[
\mbox{}
+
\bigl( x^0 \bigr)^{\cos (\phi +\theta )+\cos (\theta )}
A_{33}  \biggl(
\cot ^2(\phi ) \cos ^2(\theta )
+12 \csc ^2(\phi ) \cos (\theta )
+\frac{\csc ^2(\phi ) }{2} \bigl(\cos (2 \theta )+21\bigr)
\]
\[
\mbox{}
-2 \cot ^2(\phi ) \sin ^2(\theta )
-4 \csc (\phi ) \sin (\theta ) (4 \cos (\theta)+3)+\sin ^2(\theta )+1
\biggr)
\]
\[
\mbox{}
-8 \cot (\phi ) \csc (\phi ) 
\bigl( x^0 \bigr)^{\cos (\phi ) \cos (\theta )+\cos (\theta )+2}
A_{23}'' 
\]
\[
\mbox{}
-24 \cot (\phi ) \csc (\phi ) (\cos (\theta )+1) 
\bigl( x^0 \bigr)^{\cos (\phi ) \cos (\theta )+\cos (\theta )+1}
A_{23}' 
\]
\begin{equation}
	\mbox{}
	-16\cot (\phi ) \csc (\phi ) \cos ^2\left({\theta }/{2}\right) (\cos (\theta )+2)  
	\bigl( x^0 \bigr)^{(\cos (\phi )+1) \cos (\theta )}
	A_{23} 
	\label{EqOfSumOfDif}
	.\end{equation}

The metric of the secondary gravitational wave in the perturbative model under consideration 
for parameters values \(\mu\ne\nu\), \(\mu+\nu\ne 1\) and \(\mu,\nu\ne 1/2\) in the first approximation with respect to the small parameter \(\epsilon\) takes the following final form:
\begin{equation}
	g_{00} = 0
	,\qquad
	g_{11} = 0
	,\qquad
	g_{01} = 1-\epsilon  A_{01}
	\label{PerturbativeMetric01}
	,\end{equation}
\[
g_{02} = \epsilon\csc (\phi )\, \tau^2 \left(  B_{13} \cot (\phi )  \bigl( x^0 \bigr)^{1+\cos (\theta )}-  B_{12} \csc(\phi ) \bigl( x^0 \bigr)^{1+\sin (\phi ) \sin (\theta )+\cos (\theta )}\right)
\]
\begin{equation}
	\mbox{}
	-\epsilon  \csc ^2(\phi ) \bigl( x^0 \bigr)^{1+\cos (\theta )} \left(A_{12} \bigl( x^0 \bigr)^{\sin (\phi ) \sin (\theta )}-A_{13} \cos (\phi )\right)
	\label{PerturbativeMetric02}
	,\end{equation}
\[
g_{03} =\epsilon\csc (\phi )\, \tau^2 \left(  B_{12} \cot (\phi )  \bigl( x^0 \bigr)^{1+\cos (\theta )}-  B_{13} \csc (\phi ) \bigl( x^0 \bigr)^{1-\sin (\phi ) \sin (\theta )+\cos (\theta )}\right)
\]
\begin{equation}
	\mbox{}
	-\epsilon  \csc ^2(\phi ) \bigl( x^0 \bigr)^{1+\cos (\theta )} \left(A_{13} \bigl( x^0 \bigr)^{-\sin (\phi ) \sin (\theta )}-A_{12} \cos (\phi )\right)
	\label{PerturbativeMetric03}
	,\end{equation}
\begin{equation}
	g_{12} = -\epsilon  \csc ^2(\phi ) \bigl( x^0 \bigr)^{1+\cos (\theta )} \left(A_{02} \bigl( x^0 \bigr)^{\sin (\phi ) \sin (\theta )}-A_{03} \cos (\phi )\right)
	\label{PerturbativeMetric12}
	,\end{equation}
\begin{equation}
	g_{13} = -\epsilon  \csc ^2(\phi ) \bigl( x^0 \bigr)^{1+\cos (\theta )} \left(A_{03} \bigl( x^0 \bigr)^{-\sin (\phi ) \sin (\theta )}-A_{02} \cos (\phi )\right)
	\label{PerturbativeMetric13}
	,\end{equation}
\[
g_{22} = \csc ^2(\phi ) \bigl( x^0 \bigr)^{1+\sin (\phi ) \sin (\theta )+\cos (\theta )}
-\epsilon  \csc ^4(\phi ) \bigl( x^0 \bigr)^{2(1+ \cos (\theta ))} 
\times
\]
\begin{equation}
	\times
	\biggl[
	A_{22} \bigl( x^0 \bigr)^{2 \sin (\phi ) \sin (\theta )}
	+\cos (\phi ) \left(A_{33} \cos (\phi )
	-2 A_{23} \bigl( x^0 \bigr)^{\sin (\phi ) \sin (\theta )}\right)
	\biggr]
	\label{PerturbativeMetric22}
	,\end{equation}
\[
g_{23} = \csc ^4(\phi ) \bigl( x^0 \bigr)^{1+\cos (\theta )} 
\biggl[
\cos (\phi ) 
\,
\Bigl(
-\sin^2(\phi )
+\epsilon  A_{22} \bigl( x^0 \bigr)^{1+\sin (\phi ) \sin (\theta )+\cos (\theta )}
\]
\begin{equation}
	\mbox{}
	+\epsilon  A_{33} \bigl( x^0 \bigr)^{1-\sin (\phi ) \sin (\theta )+\cos (\theta )}
	\Bigr)
	-\epsilon  A_{23} \left(\cos ^2(\phi )+1\right) \bigl( x^0 \bigr)^{1+\cos (\theta )}
	\biggr]
	\label{PerturbativeMetric23}
	,\end{equation}
\[
g_{33} = \csc ^4(\phi ) \bigl( x^0 \bigr)^{1-2 \sin (\phi ) \sin (\theta )+\cos (\theta )} 
\,
\biggl[
\,
\sin ^2(\phi ) \bigl( x^0 \bigr)^{\sin (\phi ) \sin (\theta )}
-\epsilon  \bigl( x^0 \bigr)^{1+\cos (\theta )} 
\times
\]
\begin{equation}
	\times
	\left(\cos (\phi ) \bigl( x^0 \bigr)^{\sin (\phi ) \sin (\theta )} \
	\left(A_{22} \cos (\phi ) \bigl( x^0 \bigr)^{\sin (\phi ) \sin (\theta )}-2 A_{23}\right)+A_{33}\right)
	\biggr]
	\label{PerturbativeMetric33}
	.\end{equation}
Here \(\epsilon\) is a dimensionless small parameter (\(\epsilon\ll 1\)), 
the angular parameters \(\phi\) and~\(\theta\) are independent parameters of the wave model, the time function \(\tau(x^0,x^1,x^2,x^3)\) is defined by the relation (\ref{tauC}), 
the functions \(B_{12}(x^0)\), \(B_{13}(x^0)\) are defined 
by the equations (\ref{B12eqA})-(\ref{B13eqB}),
the functions \(A_{02}(x^0)\) and \(A_{03}(x^0)\) are defined by the equations (\ref{A02eqA})-(\ref{A03eqB}), and on the function \(A_{01}(x^0)\), \(A_{22}(x^0)\), \(A_{23}(x^0)\) 
and \(A_{33}(x^0)\) there remains only one equation (\ref{EqOfSumOfDif}) that connects them.

The determinant of the metric takes the following form:
\[
g = \det g_{\alpha\beta} = 
-
\csc ^2(\phi )\, \bigl( x^0 \bigr)^{2\bigl(1+ \cos (\theta )\bigr)}
\times
\]
\[
\times
\biggl[
1
-
\epsilon  \csc ^2(\phi ) 
\biggl(
2 A_{01} \sin ^2(\phi )+A_{22}\, \bigl( x^0 \bigr)^{\sin (\phi ) \sin (\theta )+\cos (\theta )+1}
\]
\begin{equation}
	\mbox{}
	-2 A_{23} \cos (\phi ) \bigl( x^0 \bigr)^{1+\cos (\theta )}
	+A_{33}\,  \bigl( x^0 \bigr)^{-\sin (\phi ) \sin (\theta )+\cos (\theta )+1}
	\biggr)
	\biggr]
	.\end{equation}
\[
0<\phi <\pi
,\qquad
0<\theta <\pi
.\]

The determinant \(g\) at \(x^0>0\) for 
admissible values of the parameters \(\phi\) and \(\theta\) is negative everywhere. 

Note that since \( g_{01}=g_{01}(x^0) \), then by passing to a new wave variable \( {\tilde x}^0\), such that \(d{\tilde x}^0=g_{01}(x^0)\,dx^0\), we can turn \(g_{01}\) to unity, and \(A_{01}\) to zero.
Thus, in the resulting metric of secondary gravitational waves, taking into account 
the coupling equation~(\ref{EqOfSumOfDif}), there remain two arbitrary functions of the wave variable \(x^0\).

Compared to the background wave metric, which had 3 non-zero independent components of the Riemann curvature tensor \(R_{0202}\), \(R_{0302}\) and \(R_{0303}\) in the privileged wave coordinate system, the secondary gravitational wave metric has 7 independent components of the curvature tensor. The components \(R_{0201}\), \(R_{0223}\), \(R_{0301}\), \(R_{0323}\) are added,
having the first order in \(\epsilon\).

Thus, we have obtained an analytical perturbative model of secondary gravitational waves against the background of a strong gravitational wave in the Bianchi type VI universe, where the problem is reduced to a set of ordinary differential equations whose solutions are determined by two independent angular parameters \(\phi\) and \(\theta\) of the background strong gravitational wave.

\section{On the stability of perturbative solutions}\label{sec5}

It is easy to show that there always exists a range of parameters of 
the background strong gravitational wave for which the correction functions
\(\Omega^{\alpha\beta}\) remain bounded with time \(\tau\).
Note that on geodesics (the trajectories of test bodies) in the considered metric of the background strong gravitational wave,
the time \(\tau\) is proportional to the wave variable \(x^0\) (see \cite{Osetrin325205JPA_2023}).

To simplify the field equations, we consider the case of the parameter value \(\phi=\pi/2\),
which is sufficient for the problem at hand.
Then the system of equations for the correction functions \(B_{12}\) and \(B_{13}\) (\ref{B12eqA})-(\ref{B13eqA})
is decoupled into two independent equations and simplified.
We obtain the following solutions:
\[
B_{12}=
c_{5}\left(x^0\right)^{-2(1 + \cos\theta + \sin\theta)}
,\quad
B_{13}=
c_{6}\left(x^0\right)^{-2(1 + \cos\theta - \sin\theta)}
,\quad
c_{5}, c_{6}-\mbox{const}
.\]
For these solutions, the functions \(\Omega^{12}\) and \(\Omega^{13}\)
in the range of parameters \(\phi=\pi/2\) and \(0<\theta<\pi/4\) are bounded and tend to zero with the time.

In the range of parameters \(\phi=\pi/2\), \(0<\theta<\pi/4\) in the field equations
for the correction functions \(A_{02}\) and \(A_{03}\) (\ref{A02eqB})-(\ref{A03eqB}) at large times, we can neglect 
the terms with \(B_{12}\) and \(B_{13}\) and obtain simple equations for \(A_{02}\) and \(A_{03}\), 
whose solutions in the parameter range under consideration
tend to constants over time, i.e., the corrections \(\Omega^{02}\) and \(\Omega^{03}\) are also bounded.

Finally, to determine the correction functions \(A_{22}\), \(A_{23}\), and \(A_{33}\), 
we can take three separate equations,
equating the corresponding terms in the field equation (\ref{EqOfSumOfDif}) to zero.
The solutions are functions that decrease with time.

Thus, we have shown that for the considered model of a gravitational wave in a Bianchi type VI universe,
there exists a region of parameters \(\phi\) and \(\theta\) for which all correction functions \(\Omega^{\alpha\beta}\) 
that define the perturbative model of a secondary gravitational wave remain time-bounded.
That is, the corrections do not increase with time, and the perturbative solutions are stable.

\section{A special case I of the gravitational wave for the parameter value \(\mu=1/2\) (\(\nu\ne 1/2\))}
\label{sec6}

The special case of the secondary gravity wave model considered in this section is determined by the relation for the parameter \(\mu\) of the gravity wave of the following value:
\begin{equation}
\mu = 1/2
	.\end{equation}
Then in the secondary gravity wave model there remain two parameters \(\nu\) and \(\phi\), which, due to the field equations, are related by a relation of the form:
\begin{equation}
	{\nu}^2 = \frac{(4 {\nu}+3) \cos ^2(\phi )-8 {\nu}-2}{2 (\cos (2 \phi )-3)}
	\label{MuPhiV1}
	.\end{equation}
In this case, the form of the time function \(\tau\) is determined by the relation (\ref{tauA}).

The study of the compatibility conditions of the linearized field equations for secondary gravitational waves leads to the following form of the components of the metric tensor in the first order in \(\epsilon\):
\begin{equation}
	g_{00} = 0
	,\qquad
	g_{11} = 0
	,\qquad
	g_{01} = 1-\epsilon  A_{01}({x^0})
	,\end{equation}
\[
g_{02} = \epsilon\tau^2 \left(  B_{13}({x^0}) \bigl(x^0\bigr)^{{\nu}+\frac{1}{2}} \cot (\phi ) \csc (\phi )-  {x^0} B_{12}({x^0}) \csc ^2(\phi )\right)
\]
\begin{equation}
	\mbox{}
	+\epsilon  \csc ^2(\phi ) \left(A_{13}({x^0}) \bigl(x^0\bigr)^{{\nu}+\frac{1}{2}} \cos (\phi )-{x^0} A_{12}({x^0})\right)
	,\end{equation}
\[
g_{03} = \epsilon\tau^2 \left(  B_{12}({x^0}) \bigl(x^0\bigr)^{{\nu}+\frac{1}{2}} \cot (\phi ) \csc (\phi )-  B_{13}({x^0}) \bigl(x^0\bigr)^{2 {\nu}} \csc ^2(\phi )\right)
\]
\begin{equation}
	\mbox{}
	+\epsilon  \bigl(x^0\bigr)^{{\nu}} \csc ^2(\phi ) \left(\sqrt{{x^0}} A_{12}({x^0}) \cos (\phi )-A_{13}({x^0}) \bigl(x^0\bigr)^{{\nu}}\right)
	,\end{equation}
\begin{equation}
	g_{12} = \epsilon  \csc ^2(\phi ) \left(A_{03}({x^0}) \bigl(x^0\bigr)^{{\nu}+\frac{1}{2}} \cos (\phi )-{x^0} A_{02}({x^0})\right)
	,\end{equation}
\begin{equation}
	g_{13} = \epsilon  \bigl(x^0\bigr)^{{\nu}} \csc ^2(\phi ) \left(\sqrt{{x^0}} A_{02}({x^0}) \cos (\phi )-A_{03}({x^0}) \bigl(x^0\bigr)^{{\nu}}\right)
	,\end{equation}
\[
g_{22} = -{x^0} \csc ^4(\phi ) \Bigl(
\cos ^2(\phi )-1
\]
\begin{equation}
	\mbox{}
	+
	\epsilon  {x^0} A_{22}({x^0})-2 \epsilon  A_{23}({x^0}) \bigl(x^0\bigr)^{{\nu}+\frac{1}{2}} \cos (\phi )+\epsilon  A_{33}({x^0}) \bigl(x^0\bigr)^{2 {\nu}} \cos ^2(\phi )
	\Bigr)
	,\end{equation}
\[
g_{23} = \bigl(x^0\bigr)^{{\nu}+\frac{1}{2}} \csc ^3(\phi ) 
\Bigl(
\epsilon  {x^0} A_{22}({x^0}) \cot (\phi )
+\epsilon  A_{33}({x^0}) \bigl(x^0\bigr)^{2 {\nu}} \cot (\phi )
\]
\begin{equation}
	\mbox{}
	-\sin (\phi ) \left(\epsilon  A_{23}({x^0}) \bigl(x^0\bigr)^{{\nu}+\frac{1}{2}} \left(\cot ^2(\phi )+\csc ^2(\phi )\right)+\cos (\phi )\right)
	\Bigr)
	,\end{equation}
\[
g_{33} = -\bigl(x^0\bigr)^{2 {\nu}} \csc ^4(\phi ) 
\Bigl(
\cos ^2(\phi )-1
\]
\begin{equation}
	\mbox{}
	+
	\epsilon  {x^0} A_{22}({x^0}) \cos ^2(\phi )-2 \epsilon  A_{23}({x^0}) \bigl(x^0\bigr)^{{\nu}+\frac{1}{2}} \cos (\phi )+\epsilon  A_{33}({x^0}) \bigl(x^0\bigr)^{2 {\nu}}
	\Bigr)
	.\end{equation}

The determinant of the metric tensor in the linearized approximation takes the form:
\[
g = \det g_{\alpha\beta} = \bigl(x^0\bigr)^{2 {\nu}+1} \csc ^2(\phi )
\Bigl(
-1
+
2 \epsilon A_{01}({x^0})
+\epsilon {x^0} A_{22}({x^0}) \csc ^2(\phi )
\]
\begin{equation}
	-2 \epsilon A_{23}({x^0}) \bigl(x^0\bigr)^{{\nu}+1/2} \cot (\phi ) \csc (\phi )+\epsilon A_{33}({x^0}) \bigl(x^0\bigr)^{2 {\nu}} \csc ^2(\phi )
	\Bigr)
	.\end{equation}
Note that in the metric under consideration, as in the previous case, by redefining the wave variable \(x^0\) we can turn \(A_{01}\) to zero.

The linearized field equations give a system of coupled linear homogeneous differential equations with variable coefficients on the functions \(B_{12}(x^0)\) and \(B_{13}(x^0)\) of the following form:
\[
B_{12}' = \frac{\csc ^2(\phi )\, B_{12}
}{{x^0} \left(\cos ^2(\phi )-2\right) \left((2 {\nu}-1) \log ({x^0})+4 \cos ^2(\phi )\right)}
\times
\]
\[
\times
\Bigl(
2 (2 {\nu}-1) \bigl(2 \log ({x^0})-1\bigr)
+\cos ^4(\phi ) \bigl(4 {\nu} \log ({x^0})-22 {\nu}-13\bigr)
\]
\[
\mbox{}
+\cos ^2(\phi ) \bigl((4-12 {\nu}) \log ({x^0})+14 {\nu}+9\bigr)
+(8 {\nu}+4) \cos ^6(\phi )
\Bigr)
\]
\[
+\frac{(2 {\nu}-1)\cot (\phi ) \csc (\phi ) \,\bigl(x^0\bigr)^{{\nu}-3/2}  B_{13} 
}{2 ((2 {\nu}-1) \log ({x^0})+2 \cos (2 \phi )+2)}
\times
\]
\begin{equation}
	\times
	\Bigl(
	(2 {\nu}-1) \cos (2 \phi ) \log ({x^0})+(2-4 {p}) \log ({x^0})-4
	\Bigr)
	\label{EqSysForB12V1}
	,\end{equation}
\[
B_{13}' = -\frac{
	\csc ^2(\phi )\,
	\bigl(x^0\bigr)^{-{\nu}-3/2}  
}{\left(\cos ^2(\phi )-2\right) \bigl((2 {\nu}-1) \log ({x^0})+4 \cos ^2(\phi )\bigr)}
\times
\]
\[
\times
\biggl(
\frac{\cos (\phi )\,{x^0}  B_{12} }{8}\, 
\Bigl(8 \cos (2 \phi ) (4 {\nu}-\log ({x^0})-2)+(3-6 {\nu}) \cos (4 \phi )+6 {\nu}+8 \log ({x^0})-3\Bigr)
\]
\[
\mbox{}
-2  \bigl(x^0\bigr)^{{\nu}+1/2} 
B_{13}\,
\Bigl(
-5
+
(4 {\nu}+2) \cos ^6(\phi )
+
\cos ^4(\phi ) \bigl(2 ({\nu}+1) \log ({x^0})-14 {\nu}\bigr)
\]
\begin{equation}
	+2\cos ^2(\phi ) \bigl(1- (3 {\nu}+2) \log ({x^0})+6 {\nu}\bigr)
	+2 (2 {\nu}+1) \log ({x^0})
	\Bigr)
	\biggr)
	\label{EqSysForB13V1}
	.\end{equation}

From the system of coupled equations (\ref{EqSysForB12V1})-(\ref{EqSysForB13V1}) it is possible to pass by repeated differentiation to separate equations for the functions
\(B_{12}(x^0)\) and \(B_{13}(x^0)\).
Thus, by differentiating the equation (\ref{EqSysForB12V1}) we obtain for \(B_{12}(x^0)\) a linear homogeneous differential equation with variable coefficients of the following form:
\[
B_{12}'' = \frac{
	\bigl(x^0\bigr)^{-1}B_{12}' 
}{\left(\cos ^2(\phi )-2\right) 
	\Bigl(
	2 \bigl(2 {\nu}-1\bigr) \cos ^2(\phi ) \log ({x^0})+(3-6 {\nu}) \log ({x^0})-4
	\Bigr)
}
\times
\]
\[
\times
\biggl(
-2 \cos ^4(\phi ) \bigl(10 {\nu} \log ({x^0})-2 {\nu}+\log ({x^0})+1\bigr)
+\cos ^2(\phi ) \bigl((70 {\nu}-5) \log ({x^0})-2 {\nu}+21\bigr)
\]
\[
\mbox{}
-2 \bigl(6 (5 {\nu}-1) \log ({x^0})+6 {\nu}+17\bigr)
\biggr)
\]
\[
-\frac{
	2 \bigl(x^0\bigr)^{-2}B_{12} 
}{ \left(\cos ^2(\phi )-2\right) 
	\Bigl((2 {\nu}-1) \log ({x^0})+4 \cos ^2(\phi )\Bigr)
	\Bigl(
	(2 {\nu}-1)\log ({x^0})\,
	\bigl(
	2  \cos ^2(\phi ) 
	-3 
	\bigr)-4
	\Bigr)}
\times
\]
\[
\times
\biggl(
-6
+
2 \cos ^4(\phi ) 
\bigl(
2 (6 {\nu}+1) \log ^2({x^0})
-(82 {\nu}+23) \log ({x^0})+4 ({\nu}-2)
\bigr)
\]
\[
+\cos ^2(\phi ) \bigl(-10 (6 {\nu}+1) \log ^2({x^0})+(134 {\nu}+1) \log ({x^0})+6 (8 {\nu}+7)\bigr)
\]
\begin{equation}
	\mbox{}
	+
	8 (2 {\nu}+1) \cos ^6(\phi ) \bigl(3 \log ({x^0})-1\bigr)
	+6 (6 {\nu}+1) \log ^2({x^0})
	+(14 {\nu}+5) \log ({x^0})
	\biggr)
	.\end{equation}

For \(B_{13}(x^0)\) by differentiating the equation (\ref{EqSysForB13V1}) we obtain a linear homogeneous differential equation with variable coefficients of the following form:
\[
B_{13}'' = -\frac{2\bigl(x^0\bigr)^{-1} B_{13}' 
}{(2 {\nu}-1) \left(\cos ^2(\phi )-2\right) \bigl((2 {\nu}-1) \log ({x^0})+3 \cos (2 \phi )+1\bigr)}
\times
\]
\[
\times
\biggl(
\cos ^2(\phi ) \bigl(5 (2 {\nu}+1) \log ({x^0})-70 {\nu}-7\bigr)
+
5(2 {\nu}+1)
\bigl(
3  \cos ^4(\phi )
-  \log ({x^0})
\bigr)
+20 {\nu}+2
\biggr)
\]
\[
\mbox{}
-\frac{4\left(\cos ^2(\phi )-2\right)^{-2} \bigl(x^0\bigr)^{-2} B_{13} 
}{(2 {\nu}-1)   \left((2 {\nu}-1) \log ({x^0})+4 \cos ^2(\phi )\right) ((2 {\nu}-1) \log ({x^0})+3 \cos (2 \phi )+1)}
\times\]
\[
\times
\biggl(
\cos ^6(\phi ) \bigl(2 (50 {\nu}+19) \log ({x^0})
-3 (154 {\nu}+59)\bigr)
\]
\[
\mbox{}
+\cos ^4(\phi ) 
\bigl(
4 (4 {\nu}+3) \log ^2({x^0})
-6 (54 {\nu}+13) \log ({x^0})+514 {\nu}+127
\bigr)
\]
\[
\mbox{}
-2 \cos ^2(\phi ) \left(10 (2 {\nu}+1) \log ^2({x^0})-8 (17 {\nu}+3) \log ({x^0})+70 {\nu}+13\right)
\]
\begin{equation}
	\mbox{}
	+60 (2 {\nu}+1) \cos ^8(\phi )+8 \log ({x^0}) 
	\bigl(3 {\nu} \log ({x^0})+\log ({x^0})-6 {\nu}-1\bigr)
	\biggr)
	.\end{equation}

From the linearized field equations we additionally obtain a system of coupled linear inhomogeneous differential equations for the functions \(A_{02}(x^0)\) and \(A_{03}(x^0)\):
\[
A_{02}'' = \frac{\csc ^2(\phi ) \bigl((2 {\nu}+1) \cos (2 \phi )-2\bigr) A_{02}'}{2 {x^0}}
- \Bigl( {\nu}-\frac{1}{2}\Bigr) \cot (\phi ) \csc (\phi )\, \bigl(x^0\bigr)^{{\nu}-3/2} 
A_{03}'
\]
\[
+
\frac{ 
	B_{12}
}{\left(\cos ^2(\phi )-2\right) 
	\bigl((2 {\nu}-1) \log ({x^0})+4 \cos ^2(\phi )\bigr)}
\times
\]
\[
\times
\biggl(
\cos ^2(\phi ) \bigl((4 {\nu}-6) \log ({x^0})+20 {\nu}-26\bigr)
-8 ({\nu}-1) \log ({x^0})
\]
\begin{equation}
	\mbox{}
	+(12-8 {\nu}) \cos ^4(\phi )
	-8 {\nu}+4
	\biggr)
	\label{EqSysForA02V1}
	,\end{equation}
\[
A_{03}'' =  \Bigl( {\nu}-\frac{1}{2}\Bigr) \cot (\phi ) \csc (\phi )\,   \bigl(x^0\bigr)^{-{\nu}-1/2} 
A_{02}'
+\frac{\csc ^2(\phi )  
	\bigl((2 {\nu}+1) \cos (2 \phi )-4 {\nu}\bigr)
	A_{03}'
}{2 {x^0}}
\]
\[
+\frac{
	B_{13}
}{\left(\cos ^2(\phi )-2\right) \left((2 {\nu}-1) \log ({x^0})+4 \cos ^2(\phi )\right)}
\times
\]
\[
\times
\biggl(
\cos ^2(\phi ) \bigl((4 {\nu}-6) \log ({x^0})
+20 {\nu}-26\bigr)
-8 ({\nu}-1) \log ({x^0})
\]
\begin{equation}
	\mbox{}
	+(12-8 {\nu}) \cos ^4(\phi )
	-8 {\nu}+4
	\biggr)
	\label{EqSysForA03V1}
	.\end{equation}

Note that the system of linear inhomogeneous differential equations (\ref{EqSysForA02V1})-(\ref{EqSysForA03V1}) requires that for its integration we first find the form of the functions \(B_{12}(x^0)\) and \(B_{13}(x^0)\) from the equations (\ref{EqSysForB12V1})-(\ref{EqSysForB13V1}).
Also, for the value of the parameter \(\phi= \pi/2\), the cross terms in the equations of the system (\ref{EqSysForA02V1})-(\ref{EqSysForA03V1}) due to the factor \(\cot (\phi )\) vanish and the equations of the system become independent.

From the coupled system of differential equations (\ref{EqSysForA02V1})-(\ref{EqSysForA03V1}) by repeated differentiation we can obtain separate equations for \(A_{02}(x^0)\) and \(B_{02}(x^0)\). Thus, differentiating the equation (\ref{EqSysForA02V1}), we obtain 
for the values of the parameter \(\phi\ne \pi/2\) a separate 
linear inhomogeneous differential equation with variable coefficients for the function \(A_{02}(x^0)\):
\[
A_{02}''' = -\frac{(6 {\nu}+7) A_{02}''}{2 {x^0}}
+\frac{ \left(
	12 {\nu}
	-3 (2 {\nu}+1) \cos ^2(\phi )+4
	\right)A_{02}'}{\bigl(x^0\bigr)^2 \left(\cos ^2(\phi )-2\right)}
\]
\[
+\frac{
	4\csc ^2(\phi ) \, B_{12}
}{
	{x^0} \left(\cos ^2(\phi )-2\right)^2 
	\left((2 {\nu}-1) \log ({x^0})+4 \cos ^2(\phi )\right)^2}
\times
\]
\[
\times
\biggl(
-2 \cos ^8(\phi ) \bigl((6 {\nu}-9) \log ({x^0})+36 {\nu}-50\bigr)
\]
\[
\mbox{}
+
\cos ^6(\phi ) 
\left(
(2 {\nu}-3) \log ^2({x^0})
+(42 {\nu}-57) \log ({x^0})+4 (26 {\nu}-33)
\right)
\]
\[
\mbox{}
+\cos ^4(\phi ) \left((6-4 {\nu}) \log ^2({x^0})-8 (5 {\nu}-6) \log ({x^0})-64 ({\nu}-1)\right)
\]
\[
\mbox{}
-\cos ^2(\phi ) \left((2 {\nu}+1) \log ^2({x^0})+(15-6 {\nu}) \log ({x^0})-32 {\nu}+16\right)
\]
\[
\mbox{}
+8 (2 {\nu}-3) \cos ^{10}(\phi )+2 \log ({x^0}) ((2 {\nu}-1) \log ({x^0})+2 {\nu}+3)
\biggr)
\]
\[
\mbox{}
-\frac{
	2 (2 {\nu}-1)\cot (\phi ) \csc (\phi )
	\,
	\bigl(x^0\bigr)^{{\nu}-3/2} B_{13} 
}{\left(\cos ^2(\phi )-2\right) \bigl((2 {\nu}-1) \log ({x^0})+4 \cos ^2(\phi )\bigr)^2}
\times
\]
\[
\times
\biggl(
\cos ^2(\phi ) ((3-2 {\nu}) \log ({x^0})-10 {\nu}+13)+(4 {\nu}-6) \cos ^4(\phi )+4 ({\nu}-1) \log ({x^0})+4 {\nu}-2
\biggr)
\]
\begin{equation}
	\times
	\biggl(
	\cos ^2(\phi ) \bigl((2 {\nu}-1) \log ({x^0})-2\bigr)+(2-4 {\nu}) \log ({x^0})-2
	\biggr)
	.\end{equation}

Differentiating the equation (\ref{EqSysForA03V1}), we obtain for the values of the angular parameter \(\phi\ne \pi/2\) a separate third-order differential equation for the function \(A_{03}(x^0)\):
\[
A_{03}''' = \frac{ 
	\left(20 {\nu}-5 (2 {\nu}+1) \cos ^2(\phi )\right)A_{03}''
}{(2 {\nu}-1) {x^0} \left(\cos ^2(\phi )-2\right)}
+
\frac{ \bigl(18 {\nu}-5 (2 {\nu}+1) \cos (2 \phi )+1\bigr)A_{03}'
}{(2 {\nu}-1) \bigl(x^0\bigr)^2 \left(\cos ^2(\phi )-2\right)}
\]
\[
\mbox{}
-\frac{2 (2 {\nu}-1)\cot (\phi ) \csc (\phi ) \, \bigl(x^0\bigr)^{-{\nu}-1/2}B_{12}  
}{\left(\cos ^2(\phi )-2\right) \bigl((2 {\nu}-1) \log ({x^0})+4 \cos ^2(\phi )\bigr)^2}
\Bigl((2 {\nu}-1) \log ({x^0})+5 \cos ^2(\phi )-1\Bigr)
\times
\]
\[
\times
\Bigl(
4 ({\nu}-1) \log ({x^0})
+
\cos ^2(\phi ) \bigl((3-2 {\nu}) \log ({x^0})
-10 {\nu}+13\bigr)
+(4 {\nu}-6) \cos ^4(\phi )+4 {\nu}-2
\Bigr)
\]
\[
\mbox{}
-\frac{8 B_{13}
}{(2 {\nu}-1) {x^0} \left(\cos ^2(\phi )-2\right)^2 \left((2 {\nu}-1) \log ({x^0})+4 \cos ^2(\phi )\right)^2}
\times
\]
\[
\times
\Bigl(
\cos ^4(\phi ) 
\left( 
(3-2 {\nu}) \log ^2({x^0})+(47-30 {\nu}) \log ({x^0})
-42 {\nu}+65
\right)
\]
\[
+\cos ^2(\phi ) \left((10 {\nu}-13) \log ^2({x^0})+(46 {\nu}-55) \log ({x^0})+2 {\nu}-9\right)
+4 {\nu}-2
\]
\begin{equation}
	\mbox{}
	+
	2 (2 {\nu}-3) \cos ^6(\phi ) (\log ({x^0})+5)
	+(10-8 {\nu}) \log ^2({x^0})+(6-4 {\nu}) \log ({x^0})
	\Bigr)
	.\end{equation}

Finally, from the system of linearized field equations, only one constraint equation remains for the remaining indefinite functions in the metric \(A_{22}(x^0)\), \(A_{23}(x^0)\), \(A_{33}(x^0)\) and \(A_{01}(x^0)\). Note that in this coupling equation, as in the metric, we can redefine the wave variable \(x^0\) to make \(A_{01}\) zero.

This constraint equation takes a rather cumbersome form, which, nevertheless, we will present for the sake of completeness of the presentation:
\[
0=
A_{22}({x^0}) 
\biggl(-4096  \cos ^6(\phi ) \cot ^2(\phi ) (\cos (2 \phi )-3)^6\bigl(x^0\bigr)^{3/2}
\]
\[
\mbox{}
+2048 (2 \nu -1)  \cos ^2(\phi )  (\cos (2 \phi )-3)^5\bigl(x^0\bigr)^{3/2}\log ^3({x^0})
\]
\[
\mbox{}
+12288    \cos ^4(\phi )  (\cos (2 \phi )-3)^5 \bigl(x^0\bigr)^{3/2}\log ^2({x^0})
\]
\[
\mbox{}
-1024 \sin ^2(\phi ) (\cos (2 \phi )-3)^4 \bigl(x^0\bigr)^{3/2} \log ^4({x^0}) 
\]
\[
\mbox{}
-64 (2 \nu -1)  (\cos (3 \phi )-5 \cos (\phi   ))^6 \csc ^2(\phi ) \bigl(x^0\bigr)^{3/2} \log ({x^0})\biggr)
\]
\[
\mbox{}
+ A_{23}({x^0}) \biggl(
1024 \cos ^3(\phi ) (\cos (2 \phi )-3)^5 (-14 \nu +(6 \nu +3) \cos (2 \phi )+1) \log ^3({x^0}) \bigl(x^0\bigr)^{\nu +1}
\]
\[
\mbox{}
-1024 \cos^7(\phi ) (\cos (2 \phi )-3)^6 (-18 \nu +(6 \nu +3) \cos (2 \phi )-5) \cot ^2(\phi ) \bigl(x^0\bigr)^{\nu +1}
\]
\[
\mbox{}
+96 (-18 \nu +(6 \nu +3) \cos (2 \phi )-5) (\cos (3 \phi )-5 \cos (\phi ))^5 \log ^2({x^0})  \bigl(x^0\bigr)^{\nu +1}
\]
\[
\mbox{}
-256 \cos (\phi ) (\cos (2 \phi )-3)^4 (-18 \nu +(6 \nu +3) \cos (2 \phi )-5) \log ^4({x^0}) \sin ^2(\phi ) \bigl(x^0\bigr)^{\nu +1}
\]
\[
\mbox{}
-2048 \cos ^5(\phi ) (\cos (2 \phi )-3)^6 (-14 \nu +(6   \nu +3) \cos (2 \phi )+1) 
\cot ^2(\phi ) \log ({x^0}) \bigl(x^0\bigr)^{\nu +1}\biggr)
\]
\[
\mbox{}
+ A_{33}({x^0}) \biggl(
-2048 \cos ^2(\phi ) (\cos (2 \phi )-3)^5 ((4 \nu +1) \cos (2 \phi )-6 \nu ) \log
^3({x^0}) \bigl(x^0\bigr)^{2 \nu +\frac{1}{2}}
\]
\[
\mbox{}
+1024 \cos ^6(\phi ) (\cos (2 \phi )-3)^6 ((6 \nu +5) \cos (2 \phi )-3 (6 \nu +1)) \cot ^2(\phi ) \bigl(x^0\bigr)^{2 \nu +\frac{1}{2}}
\]
\[
\mbox{}
-3072 \cos ^4(\phi ) (\cos (2   \phi )-3)^5 ((6 \nu +5) \cos (2 \phi )-3 (6 \nu +1)) \log ^2({x^0}) \bigl(x^0\bigr)^{2 \nu +\frac{1}{2}}
\]
\[
\mbox{}
+256 (\cos (2 \phi )-3)^4 ((6 \nu +5) \cos (2 \phi )-3 (6 \nu +1)) \log ^4({x^0}) \sin ^2(\phi )
\bigl(x^0\bigr)^{2 \nu +\frac{1}{2}}
\]
\[
\mbox{}
+64 ((4 \nu +1) \cos (2 \phi )-6 \nu ) (\cos (3 \phi )-5 \cos (\phi ))^6 \csc ^2(\phi ) \log ({x^0}) \bigl(x^0\bigr)^{2 \nu +\frac{1}{2}}\biggr)
\]
\[
\mbox{}
+
A_{01}'({x^0})
\biggl(
-1024 (2 \nu +1)
\bigl(x^0\bigr)^{3/2} (\cos (2 \phi )-3)^7 \cos ^8(\phi )
\]
\[
\mbox{}
+3072 (2 \nu +1) \bigl(x^0\bigr)^{3/2} (\cos (2 \phi )-3)^6 \log ^2({x^0}) \sin ^2(\phi ) \cos ^4(\phi )
\]
\[
\mbox{}
-256 (2 \nu +1) \bigl(x^0\bigr)^{3/2} (\cos (2 \phi )-3)^5   \log ^4({x^0}) \sin ^4(\phi )
\]
\[
\mbox{}
+256 \bigl(x^0\bigr)^{3/2} (\cos (2 \phi )-3)^5 (-6 \nu +(2 \nu +1) \cos (2 \phi )+1) \log ^3({x^0}) \sin ^2(2 \phi )
\]
\[
\mbox{}
-32 \bigl(x^0\bigr)^{3/2} (-6 \nu +(2 \nu +1) \cos (2 \phi )+1)   (\cos (3 \phi )-5 \cos (\phi ))^6 \log ({x^0})\biggr)  
\]
\[
\mbox{}
+
A_{22}'({x^0})
\biggl(
3072 \bigl(x^0\bigr)^{5/2} \cos ^6(\phi ) \cot ^2(\phi ) (\cos (2 \phi )-3)^7
\]
\[
\mbox{}
+3072 (2 \nu -1) \bigl(x^0\bigr)^{5/2} \cos ^4(\phi ) \cot ^2(\phi ) \log ({x^0}) (\cos (2 \phi )-3)^7
\]
\[
\mbox{}  
-1536 (2 \nu -1) \bigl(x^0\bigr)^{5/2} \cos ^2(\phi ) \log ^3({x^0}) (\cos (2 \phi )-3)^6
\]
\[
\mbox{}
-9216 \bigl(x^0\bigr)^{5/2} \cos ^4(\phi ) \log ^2({x^0}) (\cos (2   \phi )-3)^6
\]
\[
\mbox{}   
+768 \bigl(x^0\bigr)^{5/2} \log ^4({x^0}) \sin ^2(\phi ) (\cos (2 \phi )-3)^5\biggr)  
\]
\[
\mbox{}
+
A_{23}'({x^0})
\biggl(
3072 \cos ^3(\phi ) (\cos (2 \phi )-3)^5 (-6 \nu +(2 \nu +1) \cos (2 \phi )+1)
\log ^3({x^0}) \bigl(x^0\bigr)^{\nu +2}
\]
\[
\mbox{}
-3072 (2 \nu +1) \cos ^7(\phi ) (\cos (2 \phi )-3)^7 \cot ^2(\phi ) \bigl(x^0\bigr)^{\nu +2}
\]
\[
\mbox{}
+9216 (2 \nu +1) \cos ^5(\phi ) (\cos (2 \phi )-3)^6 \log ^2({x^0})
\bigl(x^0\bigr)^{\nu +2}
\]
\[
\mbox{}
-768 (2 \nu +1) \cos (\phi ) (\cos (2 \phi )-3)^5 \log ^4({x^0}) \sin ^2(\phi ) \bigl(x^0\bigr)^{\nu +2}
\]
\[
\mbox{}
-6144 \cos ^5(\phi ) (\cos (2 \phi )-3)^6 (-6 \nu +(2 \nu +1) \cos (2 \phi )+1) \cot
^2(\phi ) \log ({x^0}) \bigl(x^0\bigr)^{\nu +2}\biggr)  
\]
\[
\mbox{}
+
A_{33}'({x^0})
\biggl(
-1536 \cos ^2(\phi ) (\cos (2 \phi )-3)^5 (-6 \nu +(2 \nu +3) \cos (2 \phi )-1) \log ^3({x^0}) \bigl(x^0\bigr)^{2 \nu
	+\frac{3}{2}}
\]
\[
\mbox{} 
+6144 \nu  \cos ^6(\phi ) (\cos (2 \phi )-3)^7 \cot ^2(\phi ) \bigl(x^0\bigr)^{2 \nu +\frac{3}{2}}
\]
\[
\mbox{}
-18432 \nu  \cos ^4(\phi ) (\cos (2 \phi )-3)^6 \log ^2({x^0}) \bigl(x^0\bigr)^{2 \nu
	+\frac{3}{2}}
\]
\[
\mbox{}   
+1536 \nu  (\cos (2 \phi )-3)^5 \log ^4({x^0}) \sin ^2(\phi ) \bigl(x^0\bigr)^{2 \nu +\frac{3}{2}}
\]
\[
\mbox{}
+48 (-6 \nu +(2 \nu +3) \cos (2 \phi )-1) (\cos (3 \phi )-5 \cos (\phi ))^6 \csc ^2(\phi ) \log
({x^0}) \bigl(x^0\bigr)^{2 \nu +\frac{3}{2}}\biggr)  
\]
\[
\mbox{}
+
A_{22}''({x^0})
\biggl(
1024 \bigl(x^0\bigr)^{7/2} \cos ^6(\phi ) \cot ^2(\phi ) (\cos (2 \phi )-3)^7
\]
\[
\mbox{}
+1024 (2 \nu -1) \bigl(x^0\bigr)^{7/2} \cos ^4(\phi ) \cot
^2(\phi ) \log ({x^0}) (\cos (2 \phi )-3)^7
\]
\[
\mbox{}
-512 (2 \nu -1) \bigl(x^0\bigr)^{7/2} \cos ^2(\phi ) \log ^3({x^0}) (\cos (2 \phi )-3)^6
\]
\[
\mbox{}
-3072 \bigl(x^0\bigr)^{7/2} \cos ^4(\phi ) \log ^2({x^0}) (\cos (2 \phi   )-3)^6
\]
\[
\mbox{}
+256 \bigl(x^0\bigr)^{7/2} \log ^4({x^0}) \sin ^2(\phi ) (\cos (2 \phi )-3)^5\biggr)  
\]
\[
\mbox{}
+
A_{23}''({x^0})
\biggl(
1024 (2 \nu -1) \cos ^3(\phi ) (\cos (2 \phi )-3)^6 \log ^3({x^0}) \bigl(x^0\bigr)^{\nu
	+3}
\]
\[
\mbox{}
-2048 \cos ^7(\phi ) (\cos (2 \phi )-3)^7 \cot ^2(\phi ) \bigl(x^0\bigr)^{\nu +3}
\]
\[
\mbox{}
+6144 \cos ^5(\phi ) (\cos (2 \phi )-3)^6 \log ^2({x^0}) \bigl(x^0\bigr)^{\nu +3}
\]
\[
\mbox{}
-512 \cos (\phi ) (\cos (2 \phi )-3)^5 \log
^4({x^0}) \sin ^2(\phi ) \bigl(x^0\bigr)^{\nu +3}
\]
\[
\mbox{}
-16 (2 \nu -1) (\cos (3 \phi )-5 \cos (\phi ))^7 \csc ^2(\phi ) \log ({x^0}) \bigl(x^0\bigr)^{\nu +3}\biggr)  
\]
\[
\mbox{}
+
A_{33}''({x^0})
\biggl(
1024 \cos ^6(\phi ) (\cos (2 \phi )-3)^7 \cot ^2(\phi ) \bigl(x^0\bigr)^{2 \nu +\frac{5}{2}}
\]
\[
\mbox{}
+1024 (2 \nu -1) \cos ^4(\phi ) (\cos (2 \phi )-3)^7 \cot ^2(\phi ) 
\bigl(x^0\bigr)^{2 \nu +\frac{5}{2}}
\log ({x^0}) 
\]
\[
\mbox{}
-3072 \cos ^4(\phi ) (\cos (2 \phi   )-3)^6 \bigl(x^0\bigr)^{2 \nu +\frac{5}{2}}\log ^2({x^0}) 
\]
\[
\mbox{}
-512 (2 \nu -1) \cos ^2(\phi ) (\cos (2 \phi )-3)^6 \bigl(x^0\bigr)^{2 \nu +\frac{5}{2}}\log ^3({x^0}) 
\]
\begin{equation}
	\mbox{}
	+256 (\cos (2 \phi )-3)^5 \sin ^2(\phi ) \bigl(x^0\bigr)^{2 \nu +\frac{5}{2}}\log ^4({x^0}) 
	\biggr)  
	\label{SumFuncI}
	.\end{equation}



Note that for the special case of a gravitational wave under consideration, it is also possible to redefine the wave variable \(x^0\), which allows us to turn the metric component \(g_{01}(x^0)\) into 1, 
and the function \(A_{01}(x^0)\) into zero, both in the metric and in the field equations.
Taking into account the coupling equation (\ref{SumFuncI}) for this case of secondary gravitational waves, only two arbitrary independent functions of the wave variable  \(x^0\) remain in the model.

Thus, in the special case of a gravitational wave considered in this section for the parameter value \(\mu=1/2\) (\(\nu\ne 1/2\)) we have reduced the description of the secondary gravitational wave model to a system of ordinary differential equations with parameters \(\nu\) and \(\phi\), which satisfy the coupling equation (\ref{MuPhiV1}).

\section{Special case II of the gravitational wave for the values of the parameters
	\(\mu+\nu=1\) (\(\mu\ne\nu\)) }
\label{sec7}

The special case considered in this section is determined by the relationship between the parameters of the gravitational wave of the following form:
\begin{equation}
	\mu = 1-\nu
	.\end{equation}
Then two parameters \(\nu\) and \(\phi\) remain, which, due to the field equations, are related by a relationship of the form:
\begin{equation}
	\nu^2- \nu +\frac{1}{8}\Bigl(1+ \cos (2 \phi )\Bigr)=0
	\label{MuPhiV2}
	,\end{equation}
In this case, the form of the time function \(\tau\) is determined by the relationship (\ref{tauB}).

The study of compatibility of linearized field equations in the case under consideration yields the same dependence of the metric components on \(\tau\) as in the first case (\ref{PerturbativeMetric01})-(\ref{PerturbativeMetric33}), but the form of the equations for the functions \(A_{\alpha\beta}(x^0)\) and \(B_{\alpha\beta}(x^0)\) included in the metric will be different.

The metric in the linear approximation with respect to the parameter \(\epsilon\) takes the form:
\begin{equation}
	g_{00} = 0
	,\qquad
	g_{11} = 0
	,\qquad
	g_{01} = 1-\epsilon  A_{01}\, 
	,\end{equation}
\[
g_{02} = \epsilon\tau^2 
\left(  {x^0} B_{13}\,  \cot (\phi ) \csc (\phi )
- B_{12}\,  \bigl(x^0\bigr)^{2-2 \nu } \csc ^2(\phi )
\right)
\]
\begin{equation}
	+\epsilon  {x^0} \csc (\phi ) \left(A_{13}\,  \cot (\phi )-A_{12}\,  \bigl(x^0\bigr)^{1-2 \nu } \csc (\phi )\right)
	,\end{equation}
\[
g_{03} = \epsilon\tau^2 \left(  {x^0} B_{12}\,  \cot (\phi ) \csc (\phi )
-  B_{13}\,  \bigl(x^0\bigr)^{2 \nu } \csc ^2(\phi )
\right)
\]
\begin{equation}
	+\epsilon  \csc ^2(\phi ) \left({x^0} A_{12}\,  \cos (\phi )-A_{13}\,  \bigl(x^0\bigr)^{2 \nu }\right)
	,\end{equation}
\begin{equation}
	g_{12} = \epsilon  {x^0} \csc (\phi ) \left(A_{03}\,  \cot (\phi )-A_{02}\,  \bigl(x^0\bigr)^{1-2 \nu } \csc (\phi )\right)
	,\end{equation}
\begin{equation}
	g_{13} = \epsilon  \csc ^2(\phi ) \left({x^0} A_{02}\,  \cos (\phi )-A_{03}\,  \bigl(x^0\bigr)^{2 \nu }\right)
	,\end{equation}
\[
g_{22} = -\bigl(x^0\bigr)^{2-4 \nu } \csc ^4(\phi ) 
\biggl(
\bigl(x^0\bigr)^{2 \nu } 
\Bigl(
-1+\cos ^2(\phi )
-2 \epsilon  \cos (\phi ) {x^0} A_{23}\, 
\]
\begin{equation}
	\mbox{}
	+\epsilon  A_{33}\,  \bigl(x^0\bigr)^{2 \nu } \cos ^2(\phi )
	\Bigr)
	+\epsilon  \bigl( x^0 \bigr)^2 A_{22}\, 
	\biggr)
	,\end{equation}
\[
g_{23} = {x^0} \csc ^4(\phi ) 
\biggl(
\cos (\phi ) 
\Bigl(
-1+\cos ^2(\phi )
+
\epsilon  A_{22}\,  \bigl(x^0\bigr)^{2-2 \nu }
+\epsilon  A_{33}\,  \bigl(x^0\bigr)^{2 \nu }
\Bigr)
\]
\begin{equation}
	\mbox{}
	-\frac{1}{2}\, \epsilon  {x^0} A_{23}\,  (\cos (2 \phi )+3)
	\biggr)
	,\end{equation}
\[
g_{33} = -\csc ^4(\phi ) 
\biggl(
\bigl(x^0\bigr)^{2 \nu }
\Bigl(
-1 +\cos ^2(\phi )
-2 \epsilon  {x^0} A_{23}\,  \cos (\phi )+\epsilon  A_{33}\,  \bigl(x^0\bigr)^{2 \nu }
\Bigr)
\]
\begin{equation}
	\mbox{}
	+\epsilon  \bigl( x^0 \bigr)^2 A_{22}\,  \cos ^2(\phi )
	\biggr)
	.\end{equation}

The determinant of the metric in the linear approximation in the parameter \(\epsilon\) will take the following form:
\[
g = \det g_{\alpha\beta} = \bigl( x^0 \bigr)^2 \csc ^2(\phi )
\biggl(
-1+\epsilon \csc ^2(\phi )
\Bigl(
2 A_{01}\, \sin ^2(\phi )+A_{22}\, \bigl(x^0\bigr)^{2-2 \nu }
\]
\begin{equation}
	\mbox{}
	-2 {x^0} A_{23}\, \cos (\phi )+A_{33}\, \bigl(x^0\bigr)^{2 \nu }
	\Bigr)
	\biggr)
	.\end{equation}
Determinant metrics for \({x^0}>0\) and \(0<\phi<\pi\) is everywhere negative.

From the field equations follows a linear system of homogeneous differential equations with variable coefficients on the functions \(B_{12}(x^0)\) and \(B_{13}(x^0)\) of the following form:
\[
B_{12}' = \frac{
	B_{12} \csc ^2(\phi ) 
}{4 {x^0} \left(\cos (4 \phi ) \log ^2({x^0})-\log ^2({x^0})-8\right)}
\biggl(
80-96 \nu
\]
\[
\mbox{}
+
\cos (2 \phi ) \left(32 \nu -2 \log ^2({x^0})+\log ({x^0})-48\right)
\]
\[
\mbox{}
+2 \cos (4 \phi ) \log ({x^0}) (4 (\nu -1) \log ({x^0})+1)-8 \nu  \log ^2({x^0}) 
\]
\[
\mbox{}
+
2
\bigl(
\cos (6 \phi ) +4 
\bigr)\log ^2({x^0})
-
\bigl(
\cos (6 \phi ) +2 
\bigr)\log ({x^0})
\biggr)
\]
\[
\mbox{}
+\frac{(2 \nu -1) B_{13}\, \bigl( x^0 \bigr)^{2 (\nu -1)} \cot (\phi ) \csc (\phi ) 
}{\cos (4 \phi ) \log ^2({x^0})-\log ^2({x^0})-8}
\biggl(
12+
4
\bigl(
1
-2 \nu 
\bigr) \log ({x^0})
\]
\begin{equation}
	\mbox{}
	+
	\cos (2 \phi ) \bigl((8 \nu -4) \log ({x^0})-4\bigr)
	+
	\bigl(1
	-\cos (4 \phi ) 
	\bigr)\log ^2({x^0})
	\biggr)
	\label{EqSysForB12V2}
	,\end{equation}
\[
B_{13}' = \frac{
	\bigl( x^0 \bigr)^{-2 \nu -1} \csc ^2(\phi ) 
}{4 \left(\cos (4 \phi ) \log ^2({x^0})-\log ^2({x^0})-8\right)}
\Biggl[
4 {x^0} B_{12}({x^0}) \cos (\phi ) 
\biggl(
12-24 \nu
\]
\[
\mbox{}
+
\cos (4 \phi ) \log ({x^0}) \bigl((2 \nu -1) \log ({x^0})-1\bigr)
\]
\[
\mbox{}
+4 \cos (2 \phi ) \bigl(2 \nu +\log ({x^0})-1\bigr)
+
\bigl(
1 -2 \nu  
\bigr)\log ^2({x^0}) 
-3 \log ({x^0})
\biggr)
\]
\[
\mbox{}
+B_{13}({x^0}) \bigl( x^0 \bigr)^{2 \nu } 
\biggl(
\cos (2 \phi ) 
\left(
-32 \nu -2 \log ^2({x^0})+\log ({x^0})-16
\right)
\]
\[
\mbox{}
-2 \cos (4 \phi ) \log ({x^0}) \bigl(4 \nu  \log ({x^0})-1\bigr)
+8 \nu  \log ^2({x^0})+96 \nu +2 \cos (6 \phi ) \log ^2({x^0})
\]
\begin{equation}
	\mbox{}
	-\cos (6 \phi ) \log ({x^0})-2 \log ({x^0})-16
	\biggr)
	\Biggr]
	\label{EqSysForB13V2}
	.\end{equation}

From the system of coupled equations (\ref{EqSysForB12V2})-(\ref{EqSysForB13V2}) it is possible to pass by repeated differentiation to separate equations for the functions
\(B_{12}(x^0)\) and \(B_{13}(x^0)\).
Thus, by differentiating the equation (\ref{EqSysForB12V2}) we obtain for \(B_{12}(x^0)\) a linear homogeneous differential equation with variable coefficients of the following form:
\[
B_{12}'' = 
\frac{B_{12}'}{x^0}
\,
\Bigl(
\cos (2 \phi ) \bigl((4-8 \nu ) \log ({x^0})+4\bigr)
-\bigl(1-\cos (4 \phi ) \bigr)\log ^2({x^0})
\]
\[
+4(2\nu-1 )\log ({x^0})
-12
\Bigr)^{-1}
\times
\]
\[
\times
\Bigl(
\cos (4 \phi ) \log ({x^0}) \bigl(2 (\nu -3) \log ({x^0})+3\bigr)
+4 \cos (2 \phi ) \bigl(2 (5 \nu -3) \log ({x^0})-5\bigr)
\]
\[
+2\bigl(3- \nu   \bigr)\log ^2({x^0})
+\bigl(21 -40 \nu \bigr) \log ({x^0})
-16 \nu +68
\Bigr)
\]
\[
\mbox{}
+
\frac{B_{12}}{2\bigl(x^0\bigr)^2}
\,
\Bigl(
\log ^2({x^0})
\bigl(
1-\cos (4 \phi )
\bigr)
+8
\Bigr)^{-1} 
\times\]
\[
\times
\Bigl(
\cos (2 \phi ) \bigl((4-8 \nu ) \log ({x^0})+4\bigr)
-
\bigl(1-\cos (4 \phi ) \bigr)
\log ^2({x^0})
+
4\bigl(2 \nu  -1\bigr)
\log ({x^0})
-12
\Bigr)^{-1}
\times
\]
\[
\times
\biggl(
-4 \nu  \cos (8 \phi ) \log ^4({x^0})
-32 \nu  \cos (6 \phi ) \log ^3({x^0})
+2 \nu  \cos (8 \phi ) \log ^3({x^0})
\]
\[
+8 \cos (2 \phi ) \left((4 \nu -3) \log ^3({x^0})
+32 (\nu -1) \log ({x^0})-\log ^2({x^0})+16\right)
\]
\[
\mbox{}
+4 \cos (4 \phi ) 
\Bigl(
(4 \nu -6) \log ^4({x^0})
+(14 \nu -5) \log ^3({x^0})
+(32 \nu -55) \log ^2({x^0})
\]
\[
\mbox{}
+4 (2 \nu +5) \log ({x^0})
-12
\Bigr)
+
217\log ^2({x^0})
-512 \nu 
+688
\]
\[
\mbox{}
+
6\log ^4({x^0})
\bigl(
3 -2 \nu  
+ \cos (8 \phi ) 
\bigr)
+
\log ^3({x^0})
\bigl(
27 -58 \nu  
+24 \cos (6 \phi ) 
-7 \cos (8 \phi )
\bigr)
\]
\begin{equation}
	\mbox{}
	+
	\log ^2({x^0})
	\bigl(
	8 \cos (6 \phi ) 
	+3 \cos (8 \phi ) 
	-128 \nu 
	\bigr)
	+
	16\log ({x^0})
	\bigl(
	11
	-18\nu 
	\bigr)
	\biggr)
	.\end{equation}

For \(B_{13}(x^0)\) by differentiating the equation (\ref{EqSysForB13V2}) we obtain a linear homogeneous differential equation with variable coefficients of the following form:
\[
B_{13}'' = 
\frac{B_{13}'
}{
	4 (2 \nu -1) {x^0} 
}
\,
\Bigl(
\cos (2 \phi ) \bigl((8 \nu -4) \log ({x^0})+4\bigr)
-8 \nu  \log ({x^0})
\]
\[
\mbox{}
+\cos (4 \phi ) \log ^2({x^0})
-\log ^2({x^0})
+4 \log ({x^0})-12
\Bigr)^{-1}
\times
\]
\[
\times
\Bigl(
\mbox{}
40 \nu  \log ^2({x^0})
+8 \nu  \log ({x^0})
+480 \nu 
-224
\]
\[
\mbox{}
-\cos (2 \phi ) 
\bigl(
32 (\nu +2) \log ({x^0})
+32 (5 \nu -2)
+\log ^2({x^0})
\bigr)
\]
\[
\mbox{}
-2 \cos (4 \phi ) \log ({x^0}) 
\bigl(20 \nu  \log ({x^0})-12 \nu -9 \log ({x^0})-4\bigr)
\]
\[
\mbox{}
+\cos (6 \phi ) \log ^2({x^0})
-18 \log ^2({x^0})
+56 \log ({x^0})
\Bigr)
\]
\[
\mbox{}
-\frac{B_{13} }{8(2 \nu -1)  \bigl( x^0 \bigr)^2 
}\,
\Bigl(\bigl(\cos (4 \phi ) -1\bigr)\log ^2({x^0})-8\Bigr)^{-1} 
\times
\]
\[
\times
\Bigl(
\cos (2 \phi ) \bigl((8 \nu -4) \log ({x^0})+4\bigr)
+\bigl(\cos (4 \phi ) -1\bigr) \log ^2({x^0})
+4
\bigl(
1 -2 \nu 
\bigr) 
\log ({x^0})
-12
\Bigr)^{-1}
\times
\]
\[
\times
\biggl(
32 \nu  \cos (8 \phi ) \log ^4({x^0})
+64 \nu  \cos (6 \phi ) \log ^3({x^0})
-48 \nu  \cos (8 \phi ) \log ^3({x^0})
\]
\[
\mbox{}
+64 \nu  \cos (6 \phi ) \log ^2({x^0})
+24 \nu  \cos (8 \phi ) \log ^2({x^0})
\]
\[
\mbox{}
-2 \cos (2 \phi ) 
\Bigl(
(32 \nu +15) \log ^3({x^0})
+16 (2 \nu +1) \log ^2({x^0})
\]
\[
\mbox{}
+8 (64 \nu +1) \log ({x^0})
-512 (\nu -1)+2 \log ^4({x^0})
\Bigr)
\]
\[
\mbox{}
-8 \cos (4 \phi ) 
\Bigl(2 (8 \nu -3) \log ^4({x^0})
+(11-8 \nu ) \log ^3({x^0})
\]
\[
\mbox{}
+2 (78 \nu -31) \log ^2({x^0})
+(36-96 \nu ) \log ({x^0})
+48 \nu -24\Bigr)
\]
\[
\mbox{}
+96 \nu  \log ^4({x^0})
-16 \nu  \log ^3({x^0})
+1224 \nu  \log ^2({x^0})
+256 \nu  \log ({x^0})
+3456 \nu 
\]
\[
\mbox{}
+6 \cos (6 \phi ) \log ^4({x^0})
-12 \cos (8 \phi ) \log ^4({x^0})
-2 \cos (10 \phi ) \log ^4({x^0})
\]
\[
\mbox{}
+29 \cos (6 \phi ) \log ^3({x^0})
+6 \cos (8 \phi ) \log ^3({x^0})
+\cos (10 \phi ) \log ^3({x^0})
\]
\[
\mbox{}
+32 \cos (6 \phi ) \log ^2({x^0})
-12 \cos (8 \phi ) \log ^2({x^0})
+16 \cos (6 \phi ) \log ({x^0})
\]
\begin{equation}
	\mbox{}
	-36 \log ^4({x^0})
	+82 \log ^3({x^0})
	-484 \log ^2({x^0})
	+288 \log ({x^0})
	-1216
	\biggr)
	.\end{equation}

For the functions \(A_{02}(x^0)\) and \(A_{03}(x^0)\), included in the metric of the gravitational wave, from the linearized field equations we obtain a linear system of inhomogeneous differential equations with variable coefficients of the following form:
\[
A_{02}'' = \frac{\csc ^2(\phi )(2 \nu +\cos (2 \phi )-2) A_{02}' }{{x^0}}
-(2 \nu -1) \bigl( x^0 \bigr)^{2 (\nu -1)} \cot (\phi ) \csc (\phi ) A_{03}'
\]
\begin{equation}
	\mbox{}
	+\frac{2  \left(\cos (4 \phi ) (\log(x^0) +2) \log(x^0) -\log ^2(x^0) -2 \log (x^0) -8\right)
		B_{12}
	}{\cos (4 \phi ) \log ^2(x^0) -\log ^2(x^0) -8}
	\label{EqSysForA02V2}
	,\end{equation}
\[
A_{03}'' = (2 \nu -1) \bigl( x^0 \bigr)^{-2 \nu } \cot (\phi ) \csc (\phi ) A_{02}' +\frac{\csc ^2(\phi ) A_{03}'\,  (\cos (2 \phi )-2 \nu )}{{x^0}}
\]
\begin{equation}
	\mbox{}
	+\frac{2 B_{13}  \left(\cos (4 \phi ) (\log (x^0) +2) \log (x^0) -\log ^2(x^0) -2 \log (x^0) -8\right)}{\cos (4 \phi ) \log ^2(x^0) -\log ^2(x^0) -8}
	\label{EqSysForA03V2}
	.\end{equation}

The system of linear inhomogeneous differential equations (\ref{EqSysForA02V2})-(\ref{EqSysForA03V2}) requires for its integration to initially find the form of the functions \(B_{12}(x^0)\) and \(B_{13}(x^0)\) from the equations (\ref{EqSysForB12V2})-(\ref{EqSysForB13V2}).
For the value of the parameter \(\phi=\pi/2\), the equations in the system (\ref{EqSysForA02V2})-(\ref{EqSysForA03V2}) "disconnect" and become independent due to the presence of the multiplier \(\cot (\phi )\).

For \(\phi\ne\pi/2\) from the system of coupled equations (\ref{EqSysForA02V2})-(\ref{EqSysForA03V2}) it is also possible to pass by repeated differentiation to the equations for individual functions \(A_{02}(x^0)\) and \(A_{03}(x^0)\).
Thus, by repeated differentiation of the equation (\ref{EqSysForA02V2}) we obtain for \(A_{02}(x^0)\) a linear non-homogeneous differential equation of the third order with variable coefficients of the following form:
\[
A_{02}''' = \frac{2 (\nu -3) A_{02}''}{{x^0}}
+\frac{2(2 \nu -3) A_{02}'}{\bigl( x^0 \bigr)^2}
\]
\[
\mbox{}
+
\frac{\csc ^2(\phi )\,B_{12}
}{4 {x^0} \left(
	\log ^2({x^0})
	-\cos (4 \phi ) \log ^2({x^0})
	+8\right)^2}
\,
\biggl(
-6 \nu  \cos (6 \phi ) \log ^4({x^0})
\]
\[
\mbox{}
+12 \nu  \cos (8 \phi ) \log ^4({x^0})+2 \nu  \cos (10 \phi ) \log ^4({x^0})
-12 \nu  \cos (6 \phi ) \log ^3({x^0})
\]
\[
\mbox{}
+24 \nu  \cos (8 \phi ) \log ^3({x^0})+4 \nu  \cos (10 \phi ) \log ^3({x^0})
-32 \nu  \cos (6 \phi ) \log ^2({x^0})
\]
\[
\mbox{}
+2 \cos (2 \phi ) 
\Bigl(
2 (\nu -1) \log ^4({x^0})+(4 \nu -5) \log ^3({x^0})
\]
\[
\mbox{}
+8 (2 \nu -3) \log ^2({x^0})-24 \log ({x^0})-144
\Bigr)
\]
\[
\mbox{}
-8 \cos (4 \phi ) 
\Bigl(
(6 \nu -2) \log ^4({x^0})
+3 (4 \nu -1) \log ^3({x^0})
\]
\[
\mbox{}
+4 (14 \nu -5) \log ^2({x^0})+4 (16 \nu -5) \log ({x^0})+8
\Bigr)
\]
\[
\mbox{}
+36 \nu  \log ^4({x^0})
+72 \nu  \log ^3({x^0})+448 \nu  \log ^2({x^0})+512 \nu  \log ({x^0})
\]
\[
\mbox{}
+2048 \nu 
+6 \cos (6 \phi ) \log ^4({x^0})-4 \cos (8 \phi ) \log ^4({x^0})-2 \cos (10 \phi ) \log ^4({x^0})
\]
\[
\mbox{}
+15 \cos (6 \phi ) \log ^3({x^0})
-6 \cos (8 \phi ) \log ^3({x^0})-5 \cos (10 \phi ) \log ^3({x^0})
\]
\[
\mbox{}
+48 \cos (6 \phi ) \log ^2({x^0})+48 \cos (6 \phi ) \log ({x^0})
+32 \cos (6 \phi )-12 \log ^4({x^0})
\]
\[
\mbox{}
-18 \log ^3({x^0})
-160 \log ^2({x^0})-160 \log ({x^0})-704
\biggr)
\]
\[
\mbox{}
+\frac{
	4 (2 \nu -1)  
	\cot (\phi ) \csc (\phi )
	\bigl( x^0 \bigr)^{2 (\nu -1)}B_{13}\,
}{\left(\log ^2({x^0})-\cos (4 \phi ) \log ^2({x^0})+8\right)^2}
\times
\] 
\[
\times
\biggl(
\cos (4 \phi ) (\log ({x^0})+2) \log ({x^0})-\log ^2({x^0})
-2 \log ({x^0})-8
\biggr)
\times
\]
\[
\times
\biggl(
\cos (2 \phi ) ((4 \nu -2) \log ({x^0})-2)-4 \nu  \log ({x^0})-\cos (4 \phi ) \log ^2({x^0})
\]
\begin{equation}
	\mbox{}
	+\log ^2({x^0})+2 \log ({x^0})+10
	\biggr)
	\label{EqSysForA02PV2}
	.\end{equation}
For \(A_{03}(x^0)\) by repeated differentiation of the equation (\ref{EqSysForA03V2}) we obtain a linear non-homogeneous differential equation of the third order with variable coefficients of the following form:
\[
A_{03}''' = 
\frac{ (\cos (2 \phi )-20 \,\nu +9)A_{03}''}{2 (2 \nu -1) {x^0}}
+\frac{ (\cos (2 \phi )-8 \nu +3)A_{03}'}{(2 \nu -1) \bigl( x^0 \bigr)^2}
\]
\[
\mbox{}
+
\frac{
	4 (2 \nu -1) 
	\cot (\phi ) \csc (\phi )\,
	\bigl( x^0 \bigr)^{-2 \nu }
	B_{12} 
}{\left(-\cos (4 \phi ) \log ^2({x^0})+\log ^2({x^0})+8\right)^2}
\times
\]
\[
\times
\biggl(
\cos (4 \phi ) (\log ({x^0})+2) \log ({x^0})-\log ^2({x^0})-2 \log ({x^0})-8
\biggr)
\times
\]
\[
\times
\biggl(
\cos (2 \phi ) ((4 \nu -2) \log ({x^0})+2)-4 \nu  \log ({x^0})+\cos (4 \phi ) \log ^2({x^0})
\]
\[
-\log ^2({x^0})+2 \log ({x^0})-10
\biggr)
\]
\[
\mbox{}
+\frac{B_{13}\, \csc ^4(\phi ) 
}{64 (2 \nu -1) {x^0} \left(-\cos (4 \phi ) \log ^2({x^0})+\log ^2({x^0})+8\right)^2}
\biggl(
16384 \nu ^7 \cos (6 \phi ) \log ^4({x^0})
\]
\[
+2 \cos (4 \phi ) 
\Bigl(
\left(65536 \nu ^7+4474 \nu +2445\right) \log ^4({x^0})
-4 (51 \nu -50) \log ^3({x^0})
\]
\[
-64 (8 \nu -11) \log ^2({x^0})-128 (6 \nu -7) \log ({x^0})+256 (2 \nu -5)
\Bigr)
\]
\[
+\cos (2 \phi ) 
\Bigl(
\left(376832 \nu ^7-55376 \nu +20245\right) \log ^4({x^0})-2 (96 \nu -85) \log ^3({x^0})
\]
\[
-32 (32 \nu -41) \log ^2({x^0})-256 (6 \nu -7) \log ({x^0})+512 (25-18 \nu )
\Bigr)
+262144 \nu ^7 \log ^4({x^0})
\]
\[
+12728 \nu  \cos (6 \phi ) \log ^4({x^0})+872 \nu  \cos (8 \phi ) \log ^4({x^0})-360 \nu  \cos (10 \phi ) \log ^4({x^0})
\]
\[
+12 \nu  \cos (12 \phi ) \log ^4({x^0})+288 \nu  \cos (6 \phi ) \log ^3({x^0})+48 \nu  \cos (8 \phi ) \log ^3({x^0})
\]
\[
-96 \nu  \cos (10 \phi ) \log ^3({x^0})+24 \nu  \cos (12 \phi ) \log ^3({x^0})+1024 \nu  \cos (6 \phi ) \log ^2({x^0})
\]
\[
-256 \nu  \cos (8 \phi ) \log ^2({x^0})+1536 \nu  \cos (6 \phi ) \log ({x^0})-384 \nu  \cos (8 \phi ) \log ({x^0})
\]
\[
+1024 \nu  \cos (6 \phi )-256 \nu  \cos (8 \phi )-52840 \nu  \log ^4({x^0})+336 \nu  \log ^3({x^0})+1280 \nu  \log ^2({x^0})
\]
\[
\mbox{}
+1920 \nu  \log ({x^0})+7424 \nu -1761 \cos (6 \phi ) \log ^4({x^0})-1044 \cos (8 \phi ) \log ^4({x^0})
\]
\[
\mbox{}
-51 \cos (10 \phi ) \log ^4({x^0})+6 \cos (12 \phi ) \log ^4({x^0})-\cos (14 \phi ) \log ^4({x^0})
\]
\[
\mbox{}
-258 \cos (6 \phi ) \log ^3({x^0})
-64 \cos (8 \phi ) \log ^3({x^0})+90 \cos (10 \phi ) \log ^3({x^0})
\]
\[
\mbox{}
-16 \cos (12 \phi ) \log ^3({x^0})-2 \cos (14 \phi ) \log ^3({x^0})-1328 \cos (6 \phi ) \log ^2({x^0})
\]
\[
\mbox{}
+288 \cos (8 \phi ) \log ^2({x^0})+16 \cos (10 \phi ) \log ^2({x^0})-1792 \cos (6 \phi ) \log ({x^0})
\]
\[
\mbox{}
+448 \cos (8 \phi ) \log ({x^0})-512 \cos (6 \phi )+128 \cos (8 \phi )+14580 \log ^4({x^0})
\]
\begin{equation}
	\mbox{}
	-320 \log ^3({x^0})
	-1696 \log ^2({x^0})-2240 \log ({x^0})-9856
	\biggr)
	\label{EqSysForA03PV2}
	.\end{equation}

To integrate equations (\ref{EqSysForA02PV2})-(\ref{EqSysForA03PV2}), it is necessary to first obtain the functions \(B_{12}(x^0)\) and \(B_{12}(x^0)\) as solutions of equations (\ref{EqSysForB12V2})-(\ref{EqSysForB13V2}).

From the system of linearized field equations, only one equation remains for the remaining functions in the metric that depend on the wave variable: \(A_{22}(x^0)\), \(A_{23}(x^0)\), \(A_{33}(x^0)\) and \(A_{01}(x^0)\). This constraint equation has the following form:
\[
0=
-4 \bigl( x^0 \bigr)^{2 \nu +1} A_{01}'+2\csc ^2(\phi )\, \bigl( x^0 \bigr)^4  A_{22}''-12 (\nu -1)\csc ^2(\phi )\,  \bigl( x^0 \bigr)^3  A_{22}'
\]
\[
\mbox{}
-\csc ^2(\phi ) (12 \nu +\cos (2 \phi )-11)\, \bigl( x^0 \bigr)^2 A_{22} 
-4  \cot (\phi ) \csc (\phi )\, \bigl( x^0 \bigr)^{2 \nu +3} A_{23}''
\]
\[
\mbox{}
-12\cot (\phi ) \csc (\phi ) \bigl( x^0 \bigr)^{2 \nu +2}\!   A_{23}'
-8\cot (\phi ) \csc (\phi )  \bigl( x^0 \bigr)^{2 \nu +1}\!  A_{23} 
+
2\csc ^2(\phi ) \bigl( x^0 \bigr)^{4 \nu +2}\!   A_{33}''
\]
\begin{equation}
	\mbox{}
	+12 \nu \csc ^2(\phi )\, \bigl( x^0 \bigr)^{4 \nu +1} A_{33}'
	-\csc ^2(\phi ) (1+\cos (2 \phi )-12 \nu )\, 
	\bigl( x^0 \bigr)^{4 \nu }  A_{33}
	\label{EqOfSumOfDifII}
	.\end{equation}

Note that since in the special case under consideration in the privileged wave coordinate system the metric component
\( g_{01}\) depends only on one wave variable \(x^0\), then by redefining the variable \(x^0\) we can turn \(g_{01}\) into 1, and \(A_{01}\) into zero.
Then in the resulting metric of secondary gravitational waves,
taking into account the coupling equation (\ref{EqOfSumOfDifII}), there remain only two arbitrary functions of the wave variable \(x^0\).

Thus, in the second special case of secondary gravitational waves considered in this section for the values of the parameters of the background strong gravitational wave \(\mu+\nu=1\) (\(\mu\ne\nu\)) we also reduced the description of the model of secondary gravitational waves to a system of ordinary differential equations with the parameters of the background wave \(\nu\) and \(\phi\), which satisfy the
constraints equation (\ref{MuPhiV2}).

%
%
%

\section{Conclusion}
\label{sec8}

Based on the proper time method presented in the work, an analytical perturbative model of secondary gravitational waves against the background of strong gravitational waves in the Bianchi VI universe is constructed.
Additionally, two special cases for special values of the background wave parameters are considered. For all cases, the mathematical model of secondary gravitational waves is reduced to a system of ordinary differential equations. It is shown that there exists a continuum of gravitational wave parameters for which the perturbative solutions are stable  (do not increase with time).

The model of secondary gravitational waves in an anisotropic Bianchi VI universe constructed in the work is a rather rare case of an analytical (not numerical) perturbative model based on the exact wave solution of the Einstein equations, which can be reduced to a system of ordinary differential equations, which allows applying the approach to other similar models. The basis of the proposed approach is the use of a time variable associated with an observer who freely moves against the background of a strong gravitational wave.

The constructed model of secondary gravitational waves can be used to simulate and calculate complex phenomena at the early stages of the dynamics of the universe, including the processes of formation of inhomogeneities of dark matter, relic plasma and relic matter due to tidal accelerations and initiation of sound waves in plasma and matter. Secondary gravitational waves could have a significant effect on the formation of both the microwave electromagnetic background and the stochastic gravitational wave background of the universe. Secondary gravitational waves could also play a significant role in the processes of isotropization of the universe during its transition to the modern isotropic state.

Thus, the construction of such analytical models of secondary relic gravitational waves and their further study can clarify the processes that occurred 
in the early stages of the universe's dynamics.


\begin{thebibliography}{52}
	\providecommand{\natexlab}[1]{#1}
	\providecommand{\url}[1]{\texttt{#1}}
	\expandafter\ifx\csname urlstyle\endcsname\relax
	\providecommand{\doi}[1]{doi: #1}\else
	\providecommand{\doi}{doi: \begingroup \urlstyle{rm}\Url}\fi
	
	\bibitem[Maggiore(2000)]{MAGGIORE2000283}
	Michele Maggiore.
	\newblock Gravitational wave experiments and early universe cosmology.
	\newblock \emph{Physics Reports}, 331\penalty0 (6):\penalty0 283--367, 2000.
	\newblock ISSN 0370-1573.
	\newblock \doi{10.1016/S0370-1573(99)00102-7}.
	
	\bibitem[Ananda et~al.(2007)Ananda, Clarkson, and Wands]{PhysRevD.75.123518}
	Kishore~N. Ananda, Chris Clarkson, and David Wands.
	\newblock Cosmological gravitational wave background from primordial density
	perturbations.
	\newblock \emph{Phys. Rev. D}, 75:\penalty0 123518, Jun 2007.
	\newblock \doi{10.1103/PhysRevD.75.123518}.
	
	\bibitem[Caprini and Figueroa(2018)]{Caprini2018_163001}
	Chiara Caprini and Daniel~G. Figueroa.
	\newblock Cosmological backgrounds of gravitational waves.
	\newblock \emph{Classical and Quantum Gravity}, 35, 2018.
	\newblock \doi{10.1088/1361-6382/aac608}.
	
	\bibitem[Seljak and Zaldarriaga(1997)]{PhysRevLett.78.2054}
	Uro\u{s} Seljak and Matias Zaldarriaga.
	\newblock Signature of gravity waves in the polarization of the microwave
	background.
	\newblock \emph{Phys. Rev. Lett.}, 78:\penalty0 2054--2057, Mar 1997.
	\newblock \doi{10.1103/PhysRevLett.78.2054}.
	
	\bibitem[Bennett et~al.(2013)Bennett, Larson, Weiland, Jarosik, and Hinshaw~et
	al.]{Bennett2013}
	C.L. Bennett, D.~Larson, J.L. Weiland, N.~Jarosik, and G.~Hinshaw~et al.
	\newblock Nine-year {Wilkinson} microwave anisotropy probe {(WMAP)}
	observations: Final maps and results.
	\newblock \emph{Astrophysical Journal, Supplement Series}, 208\penalty0 (2),
	2013.
	\newblock \doi{10.1088/0067-0049/208/2/20}.
	
	\bibitem[Saito and Yokoyama(2009)]{Saito200916}
	R.~Saito and J.~Yokoyama.
	\newblock Gravitational-wave background as a probe of the primordial black-hole
	abundance.
	\newblock \emph{Physical Review Letters}, 102\penalty0 (16), 2009.
	\newblock \doi{10.1103/PhysRevLett.102.161101}.
	
	\bibitem[Saito and Yokoyama(2010)]{Saito2010867}
	R.~Saito and J.~Yokoyama.
	\newblock Gravitational-wave constraints on the abundance of primordial black
	holes.
	\newblock \emph{Progress of Theoretical Physics}, 123\penalty0 (5):\penalty0
	867--886, 2010.
	\newblock \doi{10.1143/PTP.123.867}.
	
	\bibitem[Secrest et~al.(2021)Secrest, von Hausegger, Rameez, Mohayaee, Sarkar,
	and Colin]{Secrest_2021}
	Nathan~J. Secrest, Sebastian von Hausegger, Mohamed Rameez, Roya Mohayaee,
	Subir Sarkar, and Jacques Colin.
	\newblock A test of the cosmological principle with quasars.
	\newblock \emph{The Astrophysical Journal Letters}, 908\penalty0 (2):\penalty0
	L51, feb 2021.
	\newblock \doi{10.3847/2041-8213/abdd40}.
	
	\bibitem[{Siewert, Thilo M.} et~al.(2021){Siewert, Thilo M.}, {Schmidt-Rubart,
		Matthias}, and {Schwarz, Dominik J.}]{refId0}
	{Siewert, Thilo M.}, {Schmidt-Rubart, Matthias}, and {Schwarz, Dominik J.}
	\newblock Cosmic radio dipole: Estimators and frequency dependence.
	\newblock \emph{Astronomy and Astrophysics}, 653:\penalty0 A9, 2021.
	\newblock \doi{10.1051/0004-6361/202039840}.
	
	\bibitem[Mittal et~al.(2023)Mittal, Oayda, and Lewis]{10.1093/mnras/stad3706}
	Vasudev Mittal, Oliver~T Oayda, and Geraint~F Lewis.
	\newblock {The cosmic dipole in the Quaia sample of quasars: a Bayesian
		analysis}.
	\newblock \emph{Monthly Notices of the Royal Astronomical Society},
	527\penalty0 (3):\penalty0 8497--8510, 12 2023.
	\newblock ISSN 0035-8711.
	\newblock \doi{10.1093/mnras/stad3706}.
	
	\bibitem[Ludwick and Williams(2025)]{LUDWICK2025139717}
	Kevin~J. Ludwick and Peter~L. Williams.
	\newblock Inferred {Hubble} parameter from gravitational waves in a
	perturbative {Bianchi I} background.
	\newblock \emph{Physics Letters~B}, 868:\penalty0 139717, 2025.
	\newblock ISSN 0370-2693.
	\newblock \doi{10.1016/j.physletb.2025.139717}.
	
	\bibitem[Obukhov(2024{\natexlab{a}})]{OBUKHOV2024169816}
	Valeriy~V. Obukhov.
	\newblock Classification of the non-null electrovacuum solution of
	{Einstein–Maxwell} equations with three-parameter abelian group of motions.
	\newblock \emph{Annals of Physics}, 470:\penalty0 169816, 2024{\natexlab{a}}.
	\newblock ISSN 0003-4916.
	\newblock \doi{https://doi.org/10.1016/j.aop.2024.169816}.
	
	\bibitem[Obukhov(2023{\natexlab{a}})]{ObukhovSym15030648}
	Valeriy~V. Obukhov.
	\newblock Exact solutions of {Maxwell} equations in homogeneous spaces with the
	group of motions {G3(VIII)}.
	\newblock \emph{Symmetry}, 15\penalty0 (3), 2023{\natexlab{a}}.
	\newblock ISSN 2073-8994.
	\newblock \doi{10.3390/sym15030648}.
	
	\bibitem[Nojiri and Odintsov(2007)]{Odintsov2007}
	Shin’ichi Nojiri and Sergei~D. Odintsov.
	\newblock Introduction to modified gravity and gravitational alternative for
	dark energy.
	\newblock \emph{International Journal of Geometric Methods in Modern Physics},
	04\penalty0 (01):\penalty0 115--145, 2007.
	\newblock \doi{10.1142/S0219887807001928}.
	
	\bibitem[Nojiri and Odintsov(2011)]{Odintsov2011}
	Shin’ichi Nojiri and Sergei~D. Odintsov.
	\newblock Unified cosmic history in modified gravity: {F}rom {F}({R}) theory to
	{L}orentz non-invariant models.
	\newblock \emph{Physics Reports}, 505\penalty0 (2):\penalty0 59--144, 2011.
	\newblock ISSN 0370-1573.
	\newblock \doi{10.1016/j.physrep.2011.04.001}.
	
	\bibitem[Capozziello and {De Laurentis}(2011)]{Capozziello2011}
	Salvatore Capozziello and Mariafelicia {De Laurentis}.
	\newblock Extended theories of gravity.
	\newblock \emph{Physics Reports}, 509\penalty0 (4):\penalty0 167--321, 2011.
	\newblock ISSN 0370-1573.
	\newblock \doi{10.1016/j.physrep.2011.09.003}.
	
	\bibitem[Nojiri et~al.(2017)Nojiri, Odintsov, and Oikonomou]{Odintsov2017}
	Shin’ichi Nojiri, Sergei~D. Odintsov, and Vasilis~K. Oikonomou.
	\newblock Modified gravity theories on a nutshell: {I}nflation, bounce and
	late-time evolution.
	\newblock \emph{Physics Reports}, 692:\penalty0 1--104, 2017.
	\newblock ISSN 0370-1573.
	\newblock \doi{https://doi.org/10.1016/j.physrep.2017.06.001}.
	
	\bibitem[Abbott et~al.(2017{\natexlab{a}})Abbott, Abbott, Abbott, Acernese, and
	{et al.}]{Abbott2017PRL161101}
	B.P. Abbott, R.~Abbott, T.D. Abbott, F.~Acernese, and {et al.}
	\newblock {GW170817}: Observation of gravitational waves from a binary neutron
	star inspiral.
	\newblock \emph{Physical Review Letters}, 119\penalty0 (16),
	2017{\natexlab{a}}.
	\newblock \doi{10.1103/PhysRevLett.119.161101}.
	
	\bibitem[Abbott et~al.(2017{\natexlab{b}})Abbott, Abbott, Abbott, Acernese,
	Ackley, and {et al.}]{Abbott_2017AJL}
	B.~P. Abbott, R.~Abbott, T.~D. Abbott, F.~Acernese, K.~Ackley, and {et al.}
	\newblock Gravitational waves and gamma-rays from a binary neutron star merger:
	Gw170817 and grb 170817a.
	\newblock \emph{The Astrophysical Journal Letters}, 848\penalty0 (2):\penalty0
	L13, oct 2017{\natexlab{b}}.
	\newblock \doi{10.3847/2041-8213/aa920c}.
	
	\bibitem[Abbott et~al.(2019{\natexlab{a}})Abbott, Abbott, Abbott, Acernese,
	Ackley, and {et al.}]{Abbott_PhysRevX.9.011001}
	B.~P. Abbott, R.~Abbott, T.~D. Abbott, F.~Acernese, K.~Ackley, and {et al.}
	\newblock Properties of the binary neutron star merger gw170817.
	\newblock \emph{Phys. Rev. X}, 9:\penalty0 011001, Jan 2019{\natexlab{a}}.
	\newblock \doi{10.1103/PhysRevX.9.011001}.
	
	\bibitem[Osetrin et~al.(2022{\natexlab{a}})Osetrin, Osetrin, and
	Osetrina]{Osetrin2022EPJP856}
	K.~Osetrin, E.~Osetrin, and E.~Osetrina.
	\newblock Geodesic deviation and tidal acceleration in the gravitational wave
	of the {Bianchi} type {IV} universe.
	\newblock \emph{European Physical Journal Plus}, 137\penalty0 (7),
	2022{\natexlab{a}}.
	\newblock \doi{10.1140/epjp/s13360-022-03061-3}.
	
	\bibitem[Osetrin et~al.(2022{\natexlab{b}})Osetrin, Osetrin, and
	Osetrina]{Osetrin2022894}
	K.~Osetrin, E.~Osetrin, and E.~Osetrina.
	\newblock Gravitational wave of the {Bianchi VII} universe: particle
	trajectories, geodesic deviation and tidal accelerations.
	\newblock \emph{European Physical Journal C}, 82\penalty0 (10),
	2022{\natexlab{b}}.
	\newblock \doi{10.1140/epjc/s10052-022-10852-6}.
	
	\bibitem[Osetrin et~al.(2023{\natexlab{a}})Osetrin, Osetrin, and
	Osetrina]{Osetrin325205JPA_2023}
	K~E Osetrin, E~K Osetrin, and E~I Osetrina.
	\newblock Deviation of geodesics and particle trajectories in a gravitational
	wave of the {Bianchi type~VI} universe.
	\newblock \emph{Journal of Physics A: Mathematical and Theoretical},
	56\penalty0 (32):\penalty0 325205, jul 2023{\natexlab{a}}.
	\newblock \doi{10.1088/1751-8121/ace6e3}.
	
	\bibitem[Osetrin et~al.(2023{\natexlab{b}})Osetrin, Kirnos, and
	Osetrin]{Osetrin2023Universe356}
	Konstantin Osetrin, Ilya Kirnos, and Evgeny Osetrin.
	\newblock An exact model of a gravitational wave in the {Bianchi III} universe
	based on {Shapovalov~II} wave spacetime and the quadratic theory of gravity.
	\newblock \emph{Universe}, 9\penalty0 (8), 2023{\natexlab{b}}.
	\newblock ISSN 2218-1997.
	\newblock \doi{10.3390/universe9080356}.
	
	\bibitem[Osetrin et~al.(2024{\natexlab{a}})Osetrin, Epp, and
	Chervon]{OSETRIN2024169619}
	Konstantin~E. Osetrin, Vladimir~Y. Epp, and Sergey~V. Chervon.
	\newblock Propagation of light and retarded time of radiation in a strong
	gravitational wave.
	\newblock \emph{Annals of Physics}, 462:\penalty0 169619, 2024{\natexlab{a}}.
	\newblock ISSN 0003-4916.
	\newblock \doi{https://doi.org/10.1016/j.aop.2024.169619}.
	
	\bibitem[Abbott et~al.(2016)Abbott, Abbott, Abbott, Abernathy, Acernese, and
	et~al.]{PhysRevLett.116.061102}
	B.~P. Abbott, R.~Abbott, T.~D. Abbott, M.~R. Abernathy, F.~Acernese, and et~al.
	\newblock Observation of gravitational waves from a binary black hole merger.
	\newblock \emph{Phys. Rev. Lett.}, 116:\penalty0 061102, Feb 2016.
	\newblock \doi{10.1103/PhysRevLett.116.061102}.
	
	\bibitem[Abbott et~al.(2019{\natexlab{b}})Abbott, Abbott, Abbott, Abraham,
	Acernese, and et~al.]{PhysRevX.9.031040}
	B.~P. Abbott, R.~Abbott, T.~D. Abbott, S.~Abraham, F.~Acernese, and et~al.
	\newblock {GWTC-1}: A gravitational-wave transient catalog of compact binary
	mergers observed by {LIGO} and {Virgo} during the first and second observing
	runs.
	\newblock \emph{Phys. Rev. X}, 9:\penalty0 031040, Sep 2019{\natexlab{b}}.
	\newblock \doi{10.1103/PhysRevX.9.031040}.
	
	\bibitem[Abbott et~al.(2021)Abbott, Abbott, Abraham, Acernese, Ackley, and
	et~al.]{PhysRevX.11.021053}
	R.~Abbott, T.~D. Abbott, S.~Abraham, F.~Acernese, K.~Ackley, and et~al.
	\newblock {GWTC-2}: Compact binary coalescences observed by {LIGO} and {Virgo}
	during the first half of the third observing run.
	\newblock \emph{Phys. Rev. X}, 11:\penalty0 021053, Jun 2021.
	\newblock \doi{10.1103/PhysRevX.11.021053}.
	
	\bibitem[Blanchet(2024)]{BlanchetPostNewtonian2024}
	L.~Blanchet.
	\newblock Post-newtonian theory for gravitational waves.
	\newblock \emph{Living Reviews in Relativity}, 27, 2024.
	\newblock \doi{10.1007/s41114-024-00050-z}.
	
	\bibitem[Christensen(2018)]{Christensen2018016903}
	Nelson Christensen.
	\newblock Stochastic gravitational wave backgrounds.
	\newblock \emph{Reports on Progress in Physics}, 82\penalty0 (1):\penalty0
	016903, Nov 2018.
	\newblock \doi{10.1088/1361-6633/aae6b5}.
	
	\bibitem[Osetrin and Osetrin(2020)]{Osetrin2020Symmetry}
	Konstantin Osetrin and Evgeny Osetrin.
	\newblock Shapovalov wave-like spacetimes.
	\newblock \emph{Symmetry}, 12\penalty0 (8), 2020.
	\newblock \doi{10.3390/SYM12081372}.
	
	\bibitem[Domènech(2021)]{Domenech2021398}
	G.~Domènech.
	\newblock Scalar induced gravitational waves review.
	\newblock \emph{Universe}, 7\penalty0 (11), 2021.
	\newblock \doi{10.3390/universe7110398}.
	
	\bibitem[Osetrin et~al.(2023{\natexlab{c}})Osetrin, Osetrin, and
	Osetrina]{Osetrin1455Sym_2023}
	Konstantin Osetrin, Evgeny Osetrin, and Elena Osetrina.
	\newblock Deviation of geodesics, particle trajectories and the propagation of
	radiation in gravitational waves in {Shapovalov} type {III} wave spacetimes.
	\newblock \emph{Symmetry}, 15\penalty0 (7), 2023{\natexlab{c}}.
	\newblock ISSN 2073-8994.
	\newblock \doi{10.3390/sym15071455}.
	
	\bibitem[van Remortel et~al.(2023)van Remortel, Janssens, and
	Turbang]{van_Remortel_2023}
	Nick van Remortel, Kamiel Janssens, and Kevin Turbang.
	\newblock Stochastic gravitational wave background: Methods and implications.
	\newblock \emph{Progress in Particle and Nuclear Physics}, 128:\penalty0
	104003, Jan 2023.
	\newblock \doi{10.1016/j.ppnp.2022.104003}.
	
	\bibitem[Obukhov(2025)]{Obukhov10.1142/S0219887825501774}
	V.~V. Obukhov.
	\newblock Classification of {Einstein} spaces with {Stäckel } metric of
	type~(3.0).
	\newblock \emph{International Journal of Geometric Methods in Modern Physics},
	page 2550177, 2025.
	\newblock \doi{10.1142/S0219887825501774}.
	
	\bibitem[Obukhov(2024{\natexlab{b}})]{Obukhov10.3390/sym16101385}
	V.~V. Obukhov.
	\newblock Classification of {Petrov} homogeneous spaces.
	\newblock \emph{Symmetry}, 16\penalty0 (10), 2024{\natexlab{b}}.
	\newblock ISSN 2073-8994.
	\newblock \doi{10.3390/sym16101385}.
	
	\bibitem[Osetrin et~al.(2024{\natexlab{b}})Osetrin, Epp, and
	Filippov]{Osetrin2024Symmetry1456}
	Konstantin~E. Osetrin, Vladimir~Y. Epp, and Altair~E. Filippov.
	\newblock Exact model of gravitational waves and pure radiation.
	\newblock \emph{Symmetry}, 16\penalty0 (11), 2024{\natexlab{b}}.
	\newblock ISSN 2073-8994.
	\newblock \doi{10.3390/sym16111456}.
	
	\bibitem[Osetrin et~al.(2024{\natexlab{c}})Osetrin, Filippov, and
	Osetrin]{Osetrin2024RPJ1857}
	Konstantin~E. Osetrin, Altair~E. Filippov, and Evgeny~K. Osetrin.
	\newblock Dust matter and pure radiation in a plane gravitational wave.
	\newblock \emph{Russian Physics Journal}, 67\penalty0 (11):\penalty0
	1857–1863, 2024{\natexlab{c}}.
	\newblock \doi{10.1007/s11182-024-03322-x}.
	
	\bibitem[Landau and Lifshitz(1975)]{LandauEng1}
	Lev~Davidovich Landau and Evgeniy~Mikhailovich Lifshitz.
	\newblock \emph{The Classical Theory of Fields}, volume~2 of \emph{Course of
		Theoretical Physics Series}.
	\newblock Butterworth-Heinemann, Oxford(UK), 4th edition, 1975.
	
	\bibitem[Mukhanov et~al.(1992)Mukhanov, Feldman, and
	Brandenberger]{MUKHANOV1992203}
	V.F. Mukhanov, H.A. Feldman, and R.H. Brandenberger.
	\newblock Theory of cosmological perturbations.
	\newblock \emph{Physics Reports}, 215\penalty0 (5):\penalty0 203--333, 1992.
	\newblock \doi{https://doi.org/10.1016/0370-1573(92)90044-Z}.
	
	\bibitem[Ma and Bertschinger(1995)]{Ma19957}
	C.-P. Ma and E.~Bertschinger.
	\newblock Cosmological perturbation theory in the synchronous and conformal
	{Newtonian} gauges.
	\newblock \emph{Astrophysical Journal}, 455\penalty0 (1):\penalty0 7--25, 1995.
	\newblock \doi{10.1086/176550}.
	
	\bibitem[Osetrin et~al.(2006)Osetrin, Obukhov, and Filippov]{OsetrinHomog2006}
	Konstantin~E. Osetrin, Valeriy~V. Obukhov, and Altair~E. Filippov.
	\newblock Homogeneous spacetimes and separation of variables in the
	{H}amilton--{J}acobi equation.
	\newblock \emph{Journal of Physics A: Mathematical and General}, 39\penalty0
	(21):\penalty0 6641--6647, 2006.
	\newblock \doi{10.1088/0305-4470/39/21/S64}.
	
	\bibitem[Obukhov(2022{\natexlab{a}})]{ObukhovSym14122595}
	V.~V. Obukhov.
	\newblock {Maxwell} equations in homogeneous spaces with solvable groups of
	motions.
	\newblock \emph{Symmetry}, 14\penalty0 (12), 2022{\natexlab{a}}.
	\newblock ISSN 2073-8994.
	\newblock \doi{10.3390/sym14122595}.
	
	\bibitem[Obukhov(2022{\natexlab{b}})]{ObukhovUniverse8040245}
	Valery~V. Obukhov.
	\newblock {Maxwell's} equations in homogeneous spaces for admissible
	electromagnetic fields.
	\newblock \emph{Universe}, 8\penalty0 (4), 2022{\natexlab{b}}.
	\newblock ISSN 2218-1997.
	\newblock \doi{10.3390/universe8040245}.
	
	\bibitem[Obukhov(2023{\natexlab{b}})]{Obukhov10.1063/5.0158054}
	V.~V. Obukhov.
	\newblock {{Hamilton-Jacobi} and {Klein-Gordon-Fock} equations for a charged
		test particle in space-time with simply transitive four-parameter groups of
		motions}.
	\newblock \emph{Journal of Mathematical Physics}, 64\penalty0 (9):\penalty0
	093507, 09 2023{\natexlab{b}}.
	\newblock ISSN 0022-2488.
	\newblock \doi{10.1063/5.0158054}.
	
	\bibitem[{EPTA Collaboration and InPTA Collaboration:} et~al.(2023){EPTA
		Collaboration and InPTA Collaboration:}, {Antoniadis, J.}, {Arumugam, P.},
	{Arumugam, S.}, {Babak, S.}, and {Bagchi, M. et al.}]{AandArefId0}
	{EPTA Collaboration and InPTA Collaboration:}, {Antoniadis, J.}, {Arumugam,
		P.}, {Arumugam, S.}, {Babak, S.}, and {Bagchi, M. et al.}
	\newblock The second data release from the {European} pulsar timing array -
	{III}. search for gravitational wave signals.
	\newblock \emph{Astronomy \& Astrophysics}, 678:\penalty0 A50, 2023.
	\newblock \doi{10.1051/0004-6361/202346844}.
	
	\bibitem[Reardon et~al.(2023)Reardon, Zic, Shannon, Hobbs, and {Matthew Bailes
		et al.}]{Reardon_2023}
	Daniel~J. Reardon, Andrew Zic, Ryan~M. Shannon, George~B. Hobbs, and {Matthew
		Bailes et al.}
	\newblock Search for an isotropic gravitational-wave background with the
	{Parkes} pulsar timing array.
	\newblock \emph{The Astrophysical Journal Letters}, 951\penalty0 (1):\penalty0
	L6, jun 2023.
	\newblock \doi{10.3847/2041-8213/acdd02}.
	
	\bibitem[Xu et~al.(2023)Xu, Chen, Guo, Jiang, and {Bojun Wang et al.}]{Xu_2023}
	Heng Xu, Siyuan Chen, Yanjun Guo, Jinchen Jiang, and {Bojun Wang et al.}
	\newblock Searching for the nano-hertz stochastic gravitational wave background
	with the {Chinese} pulsar timing array data release {I}.
	\newblock \emph{Research in Astronomy and Astrophysics}, 23\penalty0
	(7):\penalty0 075024, jun 2023.
	\newblock \doi{10.1088/1674-4527/acdfa5}.
	
	\bibitem[Stäckel(1897)]{Stackel1897145}
	Paul Stäckel.
	\newblock Ueber die integration der {Hamilton}'schen differentialgleichung
	mittelst separation der variabeln.
	\newblock \emph{Mathematische Annalen}, 49\penalty0 (1):\penalty0 145--147,
	1897.
	\newblock \doi{10.1007/BF01445366}.
	
	\bibitem[Shapovalov(1978{\natexlab{a}})]{Shapovalov1978I}
	Vladimir~N. Shapovalov.
	\newblock Symmetry and separation of variables in {Hamilton}-{Jacobi} equation.
	{I}.
	\newblock \emph{Soviet Physics Journal}, 21\penalty0 (9):\penalty0 1124--1129,
	1978{\natexlab{a}}.
	\newblock \doi{10.1007/BF00894559}.
	
	\bibitem[Shapovalov(1978{\natexlab{b}})]{Shapovalov1978II}
	Vladimir~N. Shapovalov.
	\newblock Symmetry and separation of variables in {Hamilton}-{Jacobi} equation.
	{I}{I}.
	\newblock \emph{Soviet Physics Journal}, 21\penalty0 (9):\penalty0 1130--1132,
	1978{\natexlab{b}}.
	\newblock \doi{10.1007/BF00894560}.
	
	\bibitem[Shapovalov(1979)]{Shapovalov1979}
	Vladimir~N. Shapovalov.
	\newblock The {St{\"{a}}ckel} spaces.
	\newblock \emph{Sib. Math. Journal (Sov. J. of Math.)}, 20\penalty0
	(5):\penalty0 790--800, 1979.
	\newblock \doi{10.1007/BF00971844}.
	
\end{thebibliography}
\end{document}